\documentclass[twocolumnm,reprint,amsmath,amssymb,aps]{revtex4-2}

\usepackage{graphicx}
\usepackage{dcolumn}
\usepackage{bm}
\usepackage{xcolor}
\usepackage[normalem]{ulem}

\usepackage[hidelinks]{hyperref}

\newcommand{\mG}{\mathcal{G}}

\newcommand{\ordo}{\mathcal{O}}
\newcommand{\vev}[1]{\left\langle #1 \right\rangle}
\newcommand{\ket}[1]{{\left|#1\right\rangle}}
\newcommand{\bra}[1]{{\left\langle #1\right|}}

\newcommand{\skalarszorzat}[2]{{\langle #1 | #2 \rangle}}

\newcommand{\qB}{\psi}
\newcommand{\Pe}{\mathcal{P}}
\newcommand{\Res}{\mathop{\rm Res}}
\newcommand{\complex}{\mathbb{C}}

\newcommand{\eps}{\varepsilon}

\newcommand{\be}[1]{\begin{equation}\label{#1}}
\newcommand{\ba}[1]{\begin{multline}\label{#1}}
\newcommand{\ee}{\end{equation}}
\newcommand{\ea}{\end{eqnarray}}

\newcommand{\so}{\scriptscriptstyle \rm A}
\newcommand{\st}{\scriptscriptstyle \rm C}
\newcommand{\sth}{\scriptscriptstyle \rm D}

\newcommand{\bu}{\bar u}

\newcommand{\bx}{\bar x}
\newcommand{\by}{\bar y}
\newcommand{\bz}{\bar z}

\newcommand{\bmu}{\bar\mu}
\newcommand{\muc}{\mu^{\scriptscriptstyle C}}
\newcommand{\mub}{\mu^{\scriptscriptstyle B}}
\newcommand{\bmuc}{\bar{\mu}^{\scriptscriptstyle C}}
\newcommand{\bmub}{\bar{\mu}^{\scriptscriptstyle B}}
\newcommand{\perm}{\mathop{\rm perm}}
\newcommand{\dire}{\mathcal{Y}}
\newcommand{\indire}{\mathcal{Z}}

\begin{document}

\preprint{APS/123-QED}

\title{Integrability breaking in the one dimensional Bose gas:\\
  Atomic losses and energy loss}

\author{A. Hutsalyuk}
 \email{hutsalyuk@gmail.com}
\author{B. Pozsgay}%
 \email{pozsgay.balazs@gmail.com}
\affiliation{%
Department of Theoretical Physics, \\
  E\"otv\"os Lor\'and University Budapest
}%
\affiliation{MTA-ELTE ``Momentum'' Integrable Quantum Dynamics Research Group,\\
  E\"otv\"os Lor\'and University Budapest}

\date{\today}

\begin{abstract}
The one dimensional $\delta$-function interacting Bose gas (the Lieb-Liniger model) is an integrable
system, which can model experiments with ultra cold atoms in one dimensional traps. 
Even though the model is integrable, integrability breaking effects are
always present  in the real world experiments. In this work we consider the integrability breaking due
to atomic loss, which is the most relevant effect in the experiments. We set up a
framework for the exact computation of the losses of the canonical charges of the model, and 
compute an exact result for the energy loss due to the local $K$-body processes, valid for arbitrary
$K$.
Our result takes the form of multiple integrals, which are explicitly factorized  in the
experimentally relevant cases of $K=1,2,3$. 
\end{abstract}

\maketitle

\section{Introduction}

One dimensional integrable models are special many body systems, whose exact solution is possible
with analytic methods. Their solvability depends on the existence of a large (typically infinite)
number of conserved quantities, which constrain the dynamical processes in the system. As an effect,
the scattering events in these models are purely elastic, and the many body $S$-matrix factorizes
into products of two body $S$-matrices \cite{Mussardo-review}. This property underlies the Bethe
Ansatz solution of these 
models \cite{Bethe-XXX}. The exact solvability typically means that the eigenstates can be constructed
analytically, nevertheless the computation of the physical observables is often quite
challenging.

The experimental advances of the last 15  
years made it possible to realize integrable systems in various cold atom experiments
(see for example
\cite{QNewtonCradle,fermi-gases-experimental-review,KiserletiOsszefogl-batchelor-foerster1}), and
this motivated the study of the non-equilibrium dynamics of integrable models
\cite{nonequilibrium-intro-review}.
A key result of the last 10 years was the understanding that the equilibration in isolated integrable models can be
described by the Generalized Gibbs Ensemble (GGE) \cite{rigol-gge,rigol-quench-review}. The GGE
involves all the conserved charges of the model, possibly including the so-called quasi-local
charges \cite{JS-CGGE,prosen-enej-quasi-local-review}.

Regarding spatially inhomogeneous situations and
quantum transport the theory of Generalized Hydrodynamics (GHD) was formulated in
\cite{jacopo-GHD,doyon-GHD}. 
Within GHD it is possible to treat both the ballistic modes and
also the diffusive corrections
\cite{doyon-jacopo-ghd-diffusive,sarang-vasseur-huse,sarang-vasseur-diff,jacopo-benjamin-bernard--ghd-diffusion-long}.
A series of recent works 
\cite{superdiff1,superdiff2,vasseur-superdiff,vir-superdiff,enej-superdiff,superuniversality} also treated the
phenomenon of super-diffusion. It is very important that 
GHD was successfully applied to describe real world experiments
\cite{ghd-experimental-atomchip,GHD-QNewton}.

One of the most interesting open problems within GHD is the treatment  of the integrability
breaking effects, which are necessarily present in the experiments.
In the strict long time limit the
integrability breaking effects completely spoil the applicability of the exact methods: the
systems eventually thermalize, or in the presence of dissipation or driving they form
non-equilibrium steady states. However, for small integrability breakind and/or for intermediate
time-scales it might be possible to  
handle these effects within GGE and GHD.

There are two main approaches to treat this problem. In the first approach the integrability breaking
terms are added as a perturbation to the Hamiltonian \cite{vasseur-breaking}. Examples include situations with slowly
varying potentials 
\cite{integr-breaking-potential} or space-time inhomogeneities in the coupling constants of the
model \cite{space-time-inhom,alvise-js-adiabatic-formation}. Weak integrability breaking and a
special class of operators that do 
not lead to thermalization on the so-called Euler scale was considered recently in \cite{doyon-weak-breaking}.

An other approach is to consider the interaction between the integrable
model and its environment. If the environment is quickly thermalizing and its response is
uncorrelated on the time scales of the integrable system, then the time evolution of its density
matrix is well described by the Lindblad equation.
It was first demonstrated in \cite{zala-tGGE} that in certain Lindblad systems a time dependent GGE can
give a very good approximation of the state of the system (see also \cite{pumping,bose-hubbard-loss}).
Within this approach the effect of long wavelength noise was studied in 
\cite{Lindblad-noise}. In contrast, the effects of localized 
Lindbladian interactions were investigated in \cite{jerome-atom-loss}.

The model considered in \cite{jerome-atom-loss} is the one dimensional $\delta$-function interacting
Bose gas (also known as the Lieb-Liniger model), which was already realized in a couple of
experiments (for a relatively recent review see the corresponding Section of \cite{kinai-LL}).
In these experiments the most relevant integrability breaking effect is that of the particle losses,
and in particular the local 3-body loss \cite{g3-rubidium,LL-g3-exp-2}
(see also \cite{bose-hubbard-loss}).
The net particle loss is given by the local 3-body correlation
function (or more generally the $K$-body correlator for the $K$-body processes), for which a number
of exact results were already computed in the literature 
\cite{zvonarev-g3,sinhG-LL1,sinhG-LL2,g3-marci-1,sajat-XXZ-to-LL,alvise-lorenzo-1,alvise-lorenzo-2}. However, 
as explained in \cite{jerome-atom-loss} (and in the closely related
work \cite{bouchoule2020breakdown}) it is also important to know the changes in the rapidity
distribution, and not only the net loss. At present there are
no exact solutions available for this problem, and  \cite{jerome-atom-loss} developed a numerical
summation method for the relevant quantities. 

In this work we set up a theoretical framework, which treats  the effect of the $K$-body losses on the
canonical conserved charges of the Lieb-Liniger model. These charges correspond to the 
moments of the rapidity distribution, and computing their time derivatives gives useful information
about the root distribution itself. As a concrete example we consider the energy loss.
The energy is the second moment of the rapidity distribution, and in parity symmetric cases it is the next
simplest quantity after the total particle number. Up to know there have been no exact results for the
energy loss in the literature.

The structure of the paper is as follows. In Section \ref{sec:lindblad} we introduce the problem and
we explain our strategy for solving it. This Section also includes
some observations and ideas that can
prove to be useful in other models as well.
In Section \ref{sec:qboson} we introduce the $q$-boson model, which is used as a lattice
regularization of the 1D Bose gas. Section \ref{sec:qloss} includes the technical computations about
the losses in the $q$-boson model. In Section \ref{sec:scaling} we perform the scaling limit towards
the Lieb-Liniger model with a finite number of particles. The thermodynamic limit is taken
afterwards in Section \ref{sec:TDL}, where the factorized formulas for the final quantities are also
presented. We discuss the results in Section \ref{sec:disc}. A number of technical computations are
relegated to the Appendices.

\section{Integrability breaking via the Lindblad equation}

\label{sec:lindblad}

In this Section we discuss integrability breaking effects in general, without specifying the concrete
model. We discuss the key concepts such as the Lindblad approximation, the GGE and
the so-called string charge relations in Bethe Ansatz solvable models.
The aim of this Section is to highlight the general ideas behind our computations. Sections
\ref{sec:gge}-\ref{sec:lindbladstrat} discuss a standard strategy which can be applied in a number
of integrable models, such as the Heisenberg spin chains. However, our concrete
model, the 1D Bose gas is rather special, and some of the standard methods do not work in this
case.
This is explained in Section \ref{sec:liebstrat}, where we also highlight our strategy specially
adapted to the Lieb-Liniger model.

\subsection{The Lindblad equation}

Let $\varrho$ be the density matrix of our system. For simplicity let
us assume here that our model has finite number of degrees of freedom, and let
$H$ be the Hamiltonian. Systems with continuous degrees of freedom (such as our main model,  the interacting Bose
gas) will be considered later.

We assume that our system is in contact with a Markovian environment:  this means that the response
of the environment is uncorrelated on the time scales
of our model. In this case the evolution of the reduced density matrix of the system is well approximated by the
Lindblad equation \cite{lindblad-eredeti,lindblad-intro}:
\begin{equation}
  \dot\varrho=-i[H,\varrho]+\sum_{a} \gamma_a
\left(L_a \varrho L_a^\dagger-\frac{1}{2}\left\{L_a^\dagger L_a,\varrho \right\}\right).
\end{equation}
Here the $L_a$ are the so-called Lindblad or jump operators, that
describe concrete processes of the system, mediated by the
environment. The $\gamma_a$ are non-negative coupling constants.

Within the Lindblad approach the time dependence of any quantity
$\ordo$ is given by
\begin{multline}
  \label{lindblad2}
\frac{d}{dt}  \vev{\ordo}=\text{Tr}(\ordo \dot \varrho)=-i\vev{[\ordo,H]}\\
+\sum_{a} \gamma_a
\left(\vev{L^\dagger_a \ordo L_a}-\frac{1}{2}\vev{\{\ordo,L_a^\dagger L_a\}}\right).
\end{multline}

It is our aim to compute the effect of the Lindblad terms on the equilibrated steady states of the
integrable models. To this order first we discuss the nature of the equilibrium states.

\subsection{GGE and the string-charge relations}

\label{sec:gge}

Integrable models possess a set of conserved charges $\{Q_\alpha\}_{\alpha=1,2,\dots}$, which
commute with each other, and the Hamiltonian is a member of the
series. The construction of these charges can depend on the concrete model, but for a large class of
systems they are obtained from a commuting set of transfer matrices (TM's) \cite{Korepin-Book}.

The extra conservation laws restrict the possible dynamical processes in the system: they only allow
elastic and completely factorized scattering events \cite{Mussardo-review,caux-integrability}. As an
effect, the equilibrated steady states and the transport properties of these models are markedly
different from a generic quantum mechanical model. Regarding equilibration it is now understood that
in integrable systems which are sufficiently well isolated from the environment the steady states
can be described by the Generalized Gibbs Ensemble (GGE). The idea behind the GGE is to involve all
conserved charges of the model, thus a GGE density matrix has the form
\begin{equation}
  \varrho=\frac{e^{-\sum_j \beta_jQ_j}}{Z},\qquad Z=\text{Tr}\ e^{-\sum_j \beta_jQ_j}.
\end{equation}
Originally the GGE was devised for free models \cite{rigol-gge,rigol-2}, where the charges can be
chosen as the mode occupation numbers. In these cases the $\beta_j$ parameters can be interpreted as mode
dependent inverse temperatures. In interacting models the situation is more complicated, and the
GGE density matrix was an object of active interest for a couple of years. In the prototypical
example of the Heisenberg spin chain it was first understood that it is not enough to include the
canonical local charges \cite{JS-oTBA,sajat-oTBA} and a complete GGE requires also the recently
discovered quasi-local charges \cite{JS-CGGE,prosen-enej-quasi-local-review}. 
Furthermore it was understood that the GGE is basically equivalent to specifying
the so-called Bethe root densities \cite{JS-CGGE,jacopo-massless-1,enej-gge}; this
is known as the string-charge duality. Now we describe this
connection, which is the basis of our computations.

In Bethe Ansatz solvable models the finite volume eigenstates are
characterized by a set of Bethe roots, which describe the momenta of
the particles within the interacting multi-particle state. In the
thermodynamic limit (TDL) the states can be described by the root distribution
functions $\rho_s(\lambda)$, where $\lambda$ is the rapidity parameter
and the discrete index $s$ stands for the various particle types that
exist in the model (for a precise definition see Section \ref{sec:TDL} below). It is generally understood that in the TDL
the root densities completely determine the correlation functions, thus the root densities carry all
relevant information about the equilibrated states.
In accordance, a set of charges is complete for the construction of a GGE, if the set of their eigenvalues 
completely specifies all Bethe root densities
\cite{JS-oTBA,sajat-oTBA,sajat-GETH,JS-CGGE,jacopo-massless-1,enej-gge}.

The charges in question are extensive and their eigenvalues are additive. In the TDL the eigenvalues
are typically expressed as
\begin{equation}
  \Lambda_\alpha=\int d\lambda\ \mathfrak h_\alpha(\lambda) \rho(\lambda),
\end{equation}
where $\mathfrak h_\alpha(\lambda)$ is the one-particle eigenvalue
function.  For simplicity we assumed
here that there is only one particle species in the spectrum. If the fundamental particles can form
bound states (which is the case for the so-called string solutions in the Heisenberg chains or the
attractive Lieb-Liniger model), then all the bound states have to be treated as different particles,
and the above formula needs to be supplemented with a summation over the particle species.

It is convenient to introduce
generating functions for the charges. For example let us define
\be{charges_gen}
  X(u)=\sum_{\alpha=2}^\infty \frac{u^{\alpha-2}}{(\alpha-2)!}Q_\alpha.
\ee
The eigenvalues of this operator are given by
\be{Lau}
\Lambda(u)=\int d\lambda\ \mathfrak h(u,\lambda) \rho(\lambda),
\end{equation}
where
\be{h-generator}
\mathfrak  h(u,\lambda) =\sum_{\alpha=2}^\infty
\frac{u^{\alpha-2}}{(\alpha-2)!} \mathfrak h_\alpha(\lambda).
\ee
Completeness of the charges means that the eigenvalue function $\Lambda(u)$ completely specifies the root
density. In other words, a set of charges is complete, if the integral transform \eqref{Lau} can be
inverted. The specific details of the formulas of the type \eqref{Lau} depend on the model and its
particle content; in the Heisenberg chain the existence of the relations was called {\it
  string-charge duality} in \cite{jacopo-massless-1}. 

There are a few concrete models where the relation \eqref{Lau} and its inversion is
established. The most prominent example is the XXZ Heisenberg spin chain. That model allows a large
variety of string solutions (bound 
states of spin waves) depending on the anisotropy parameter. In that model the relation \eqref{Lau}
needs to be generalized to include all string types and all quasi-local charges. The upshot is
that eventually the integral transforms are easily inverted (for details see
\cite{JS-CGGE,jacopo-massless-1,enej-gge}). A more simple situation is that of the so-called
$q$-boson lattice model, where the relation \eqref{Lau} is
basically just a Fourier transform \cite{sajat-qboson} (see later in Section \ref{sec:qboson}).

The situation in the continuum Bose gas is more problematic, and it will be discussed in Section
\ref{sec:liebstrat}. 

\subsection{String-charge relations and the Lindblad equation}

\label{sec:lindbladstrat}

Our aim is to compute the changes in the Bethe root distributions, as an effect of the Lindblad jump
operators. The most natural idea is to compute the time derivative of the charges via relation
\eqref{lindblad2}, and then to use the 
inversion of the relation \eqref{Lau} to find the time derivative of the root densities. 

If the system is equilibrated at time $t$, then its density matrix commutes with the charges. In
this case the time derivative of the charge generating function is given by
\be{Xderiv}
\frac{d}{dt}  \vev{X(u)}=
\sum_{a}\gamma_a\vev{L^\dagger_a [X(u),L_a]}.
\ee
The remaining step is the exact computation of the r.h.s. above, and the application of relation \eqref{Lau}.

In integrable models the standard framework to treat objects like the r.h.s. above is the Algebraic Bethe
Ansatz (ABA) \cite{Korepin-Book}. This method naturally accommodates the conserved charges, and also
the local operators of the models. Furthermore, commutation relations and mean values of operator
products are relatively
easily derived, thus ABA is a promising choice for the computation of \eqref{Xderiv}.

In ABA the charges are derived from a commuting set of transfer matrices $t(u)$. These transfer
matrices are constructed from local Lax operators; the specific construction will be given below. For
the moment let us just remind that the generating functions $X(u)$ are typically defined through
relations like \cite{JS-CGGE,prosen-enej-quasi-local-review} 
\be{Xudef}
  X(u)=\partial_\lambda \log(t(u))=t^{-1}(u) t'(u).
\ee
Such a definition ensures that the $X(u)$ are extensive, and their mean values are additive.

In ABA it is relatively easy to compute the action of the transfer matrices, or commutation
relations with them. However, the definition \eqref{Xudef} also involves the inverse of the 
transfer matrix, and generally it is not known how
to handle that operator within the ABA. 

One possibility is to insert a complete set of states; such a step was also used in the numerical
summation scheme of \cite{jerome-atom-loss}. An other possibility is to 
use an {\it asymptotic inverse} of the transfer matrix, which exists in a number of models including
the Heisenberg chains. 

Let us therefore assume that there exists an other transfer matrix $\bar t(u)$,
which commutes with the original TM, and which satisfies the
asymptotic inversion relation
\be{asymp}
  \bar t(u) t(u)=1+\ordo(e^{-\xi L}).
\ee
This relation should hold at least in some neighborhood of $u=0$, the exponent $\xi$ can depend on
$u$, but it is required that $\xi(0)>0$. If  these conditions hold then it is safe to substitute $\bar
t(u)$ for $t^{-1}(u)$ in \eqref{Xudef}, and for the time derivatives in the TDL we obtain
\be{Xderiv2}
\frac{d}{dt}  \vev{X(u)}=
\sum_{a}\gamma_a\vev{L^\dagger_a [\bar t(u) t'(u),L_a]}+\dots,
\ee
where the dots denote exponentially small corrections that scale to zero in the TDL. Computation of
the r.h.s. of \eqref{Xderiv2} is a relatively standard task in ABA, and together with the inversion
of \eqref{Lau} this can be considered the ``canonical'' way of approaching integrability breaking in
ABA.

An asymptotic inverse of the transfer matrix exist in many models, including the
Heisenberg spin chains and their higher rank generalizations
\cite{essler-xxz-gge,prosen-xxx-quasi,sajat-su3-gge}; in the XXZ chain
the asymptotic inverse is simply the original transfer matrix
evaluated at a shifted rapidity \cite{essler-xxz-gge,prosen-xxx-quasi}.
However, such inversion relations seem to exist only in those models
where the fundamental particles can form bound states \cite{JS-CGGE,jacopo-massless-1}.
As far as we know there is no asymptotic inverse for the TM 
of the repulsive Bose gas. Thus the ``canonical'' strategy of ABA does not seem to be applicable in this
model, and alternative ways are needed. This is discussed in the Subsection below.

\subsection{The repulsive Lieb-Liniger model}

\label{sec:liebstrat}
The Lieb-Liniger model is a continuum theory which is given in second quantized formalism as
\be{LiebLiniger}
  H=\int_0^L dx \left[\partial_x\Psi^\dagger(x)\partial_x \Psi(x)+c \Psi^{2\dagger}(x)\Psi^2(x)\right].
\ee
Here $\Psi(x)$ is a canonical Bose field. The number $c$ is the coupling constant of the model, and
we treat here the repulsive case with $c>0$.

The model is solved by the Bethe Ansatz \cite{Lieb-Liniger}. The interacting multi-particle states
can be characterized by a set of rapidities $\{p_1,\dots,p_N\}$, and the (un-normalized) wave functions
are given explicitly as
    \begin{multline}
      \label{egyfajta-coo}
  \chi_N(\{p\}|\{x\})=\frac{1}{\sqrt{N!}} \sum_{\Pe\in S_N}
  e^{ i\sum_j x_j (\Pe p)_j}
  \\ \times \prod_{j>k}
\frac{(\Pe p)_j-(\Pe p)_k-ic\epsilon(x_j-x_k)}{(\Pe p)_j-(\Pe p)_k}.   
   \end{multline}
Here the summation runs over all permutations $\Pe\in S_n$ and $\epsilon(x)$
is the sign function.

Periodic boundary conditions imply that the rapidities satisfy
the Bethe Ansatz equations
\be{LL-BA-e}
  e^{ip_jL}\prod_{k\ne j} \frac{p_j-p_k-ic}{p_j-p_k+ic}=1.
\ee
The energy and momentum of the multi-particle state is given by
\begin{equation*}
  E=\sum_j p_j^2\qquad\qquad P=\sum_j p_j.
\end{equation*}
Generally it is assumed that  there are higher canonical charges
$Q_\alpha$ with their finite volume eigenvalues being
\be{Laalpha}
  \Lambda_\alpha=\sum_j p_j^\alpha.
\ee
In this notation we have $E=Q_2$ and $P=Q_1$.

These canonical charges can be found from a transfer matrix construction, appropriate for the
continuous space  \cite{Korepin-Book}.
However, the expansion of the transfer matrix into a
discrete set of charges gives singular operators. 
This was discussed in detail in \cite{korepin-LL-higher}. It was found there, that the operator expressions
for the higher charges (starting from $Q_4$) contain a number of singular terms (for example Dirac
delta squared), which can not be immediately canceled. However, the action of the charges remains
finite on the eigenstates.

The reason for the apparent singularity of the higher operators lies in the special form of the
Bethe wave function.
It can be seen from \eqref{egyfajta-coo} that the wave function is continuous in all of its
variables, but due to the interaction 
there are jumps in its space derivatives whenever $x_j=x_k$ for some $j,k$. It follows that the
higher space derivatives can not be defined at all. The higher conserved charges naturally
involve higher space derivatives, and this incompatibility with the wave function underlies the
singularities discussed in 
\cite{korepin-LL-higher}.
On the other hand, a lattice regularization yields finite and well defined action of the charges on
the eigenstates.

It follows from \eqref{Laalpha} that in the TDL the densities of the charge eigenvalues are
\begin{equation}
\frac{\Lambda_\alpha}{L}=\int dp\  p^\alpha \rho(p),
\end{equation}
where $\rho$ is the density of the rapidities $\{p_k\}$ (for precise definition see Section
\ref{sec:TDL}). In other words the charges measure the moments of the root distribution. If all the $\Lambda_\alpha$ are
known, then in principle $\rho(p)$ can be reconstructed, although this might not be practical in
concrete applications.

Let us now discuss the atomic losses within the Lindblad approach. In this model the experimentally relevant
processes are the local $K$-body losses, the main contribution being the 3-body loss
\cite{g3-rubidium,LL-g3-exp-2}. The discrete Lindblad equation has to be replaced by a continuous
version:
\begin{multline}
  \dot\varrho=-i[H,\varrho]\\+G \int dx 
\left(L(x) \varrho L^\dagger(x)-\frac{1}{2}\left\{L^\dagger(x) L(x),\varrho \right\}\right),
\end{multline}
where now $G$ is an overall coupling constant.
The Lindblad jump operators are
\be{LLLindblad}
  L(x)=\Psi^K(x),\qquad  L^\dagger(x)=\Psi^{\dagger K}(x).
\ee
We should compute the time derivative of the Bethe root densities under this Lindbladian time
evolution. As explained above, we approach the problem by first looking at the canonical charges. 
Thus we intend to calculate
\be{Xderiv3}
\frac{d}{dt}  \vev{Q_\alpha}= G\int dx
\vev{L^\dagger(x) [Q_\alpha,L(x)]}.
\ee
Assuming spatially homogeneous situations the time derivative of the density is found simply from a
local action of the jump operators:
\be{Xderiv3b}
{\frac{d}{dt} \frac{\vev{Q_\alpha}}{L}= G
\vev{L^\dagger(0) [Q_\alpha,L(0)]}.}
\ee
However, it is not immediately clear how to proceed from here. In principle the charges can be
computed from a transfer matrix using the appropriate version of relations \eqref{charges_gen}-\eqref{Xudef}
\cite{Korepin-Book,korepin-LL-higher}, but there is no asymptotic inverse as far as we know. Thus it
is not clear how to apply the ABA formalism to compute the r.h.s. above. 

Instead of the ambitious goal of solving \eqref{Xderiv3} for all $\alpha$
let us
focus on the simplest charges. This can already give useful information, and it might be a starting
point for the general case.

The simplest charge in the series is the net particle number
\be{partNumber}
  N=Q_0=\int dx\ \Psi^\dagger(x)\Psi(x).
\ee
Its time derivative in equilibrium is given by
\be{Nderiv}
  \frac{d}{dt}  \frac{\vev{N}}{L}=
  G 
\vev{L^\dagger(0) [N,L(0)]}=
-GKg_K,
\ee
where
\be{gK-def}
  g_K=\vev{\Psi^{\dagger K}(0)\Psi^K(0)}
\ee
is the local $K$-body correlation function. In deriving \eqref{Nderiv} we just used the commutation
relations of the field operators.

The local correlation functions have been an object of active interest, and a number of exact results
for $g_K$ were computed in the literature
\cite{zvonarev-g3,sinhG-LL1,sinhG-LL2,g3-marci-1,sajat-XXZ-to-LL,alvise-lorenzo-1,alvise-lorenzo-2}. Remarkably
none of these works dealt directly with the Lieb-Liniger model; instead they considered various
scaling limits of other models. The works \cite{zvonarev-g3,sajat-XXZ-to-LL} used the $q$-boson
model and the XXZ Heisenberg chain as
lattice regularizations, whereas
\cite{sinhG-LL1,sinhG-LL2,g3-marci-1,alvise-lorenzo-1,alvise-lorenzo-2} used a special
non-relativistic limit of the sinh-Gordon model.

Let us now turn to the next simplest charge. 
If the initial Bethe root distribution is parity symmetric, then the overall momentum is zero and
the next relevant charge is 
the energy. We intend to compute
\be{Hlossrate}
\frac{d}{dt}  \frac{\vev{H}}{L}= G 
  \vev{\Psi^{\dagger K}(0) [H,\Psi^K(0)]  }.
\ee
To gain some insight to the problem, we can attempt a direct evaluation of the commutator above. However,
this yields singular contributions. For example, a formal substitution of  the interaction term of $H$ into the r.h.s. gives 
\be{d1}
  \begin{split}
-GcK\int dx\  & \delta(x) \left[ \Psi^{\dagger K}(0)\Psi^\dagger(x)  \Psi^{K-1}(0) \Psi^2(x)\right.\\
&+\left.  \Psi^{\dagger K}(0) \Psi^{K-1}(0)  \Psi^\dagger(x) \Psi^2(x)\right].    
  \end{split}
\ee
This expression can not be normal ordered, and we obtain a singular contribution. 
A different singular term is also obtained as we substitute the kinetic term in $H$.
Altogether we observe a situation similar to that of the higher conserved charges discussed in
\cite{korepin-LL-higher}: we get singular contributions on the operator level, but we expect that their
action on the Bethe states remains finite.
In order to unambiguously obtain these finite terms a lattice regularization has to be
applied. 
However, before turning to the lattice model let us compute the finite term in 
\eqref{d1}. Dropping the singular term and taking the mean value we get
\be{GcK}
 - 2GcK  \vev{ \Psi^{\dagger (K+1)}(0)  \Psi^{K+1}(0) }=-2GcK g_{K+1}.
\ee
This means that in the final exact result there should be a term proportional to the local
$K+1$-body correlation function. Below we show that this is indeed the case, and we obtain the
proportionality factors as given above.

Regarding the lattice discretization we choose the so-called $q$-boson model, which is introduced in
the next Section. Two important advantages of the $q$-boson model are that
the string-charge relations are simple (they consist of a mere Fourier
transform, see below), and that the scaling limit towards the Lieb-Liniger model is rather straightforward.

To close this Section we give more comments on the different regularization schemes. The singular
terms obtained above do not seem to depend on the states, and it might be possible to subtract them
directly in the continuum. However, the main advantage of the lattice regularization is that it yields all the
higher charges and their action on the Bethe states. On the contrary, the regularization in the
continuum would necessarily become more complicated, as already shown by the charges themselves
\cite{korepin-LL-higher}.

\section{The $q$-boson model and its scaling limit}

\label{sec:qboson}

The $q$-boson model is a model of interacting bosons on the lattice, originally constructed in
\cite{qbozon-bog-bullough1,qbozon-bog-bullough2,qbozon-bog-bullough3} and 
further analyzed in \cite{qboson-izergin-kitanine-bog}.
 It  can serve as a lattice regularization of the Lieb-Liniger model \cite{zvonarev-g3}, and
it has connections to combinatorics \cite{qboson-bog1,qboson-bog2,qboson-bog3,qboson-keiichi}
and the theory of symmetric functions \cite{qboson-symmetric}.
In this Section we introduce the
model and its Algebraic Bethe Ansatz solution, following the conventions of \cite{qboson-izergin-kitanine-bog}.

Consider a lattice consisting of $L$ sites such that the
configuration space of each site is a single bosonic space. We define the canonical Bose operators
$a_j$, $a^\dagger_j$, and $N_j$ acting on site $j$ by the usual commutation relations
\begin{equation*}
  [a_j,a_k^\dagger]=\delta_{j,k},
\end{equation*}
\begin{equation*}
[N_j,a_k]=-\delta_{j,k}  a_k,\qquad
[N_j,a_k^\dagger]=\delta_{j,k}  a_k^\dagger.
\end{equation*}
The action of these operators on the local states $\ket{n}_j$,
$n=0\dots\infty$ is given by
\begin{equation*}
  a_j\ket{n}_j=\sqrt{n}\ket{n-1}_j,\qquad
 a_j^\dagger\ket{n}_j=\sqrt{n+1}\ket{n+1}_j,
\end{equation*}
\begin{equation*}
 N_j\ket{n}_j=n\ket{n-1}_j.
\end{equation*}
Let us also define the local $q$-boson field operators $\qB_j^\dagger$, $\qB_j$ through their action
\begin{equation*}
  \qB_j\ket{n}_j=\sqrt{[n]_q}\ket{n-1}_j\qquad
 \qB_j^\dagger\ket{n}_j=\sqrt{[n+1]_q}\ket{n+1}_j,
\end{equation*}
where
\begin{equation*}
  [x]_q=\frac{1-q^{-2x}}{1-q^{-2}}.
\end{equation*}
The parameter $q$ is a real number which is related to the interaction strength in the model.
We will consider the regime $q\ge 1$ and we will use the parameterization
$q=e^{\eta}$, $\eta>0$.

These $q$-deformed operators satisfy the following
commutation relations:
\begin{equation*}
[N_k,\qB_k]=-  \qB_k,\quad
{[N_k,\qB_k^\dagger]=\qB_k^\dagger,}\quad
[\qB_k,\qB_k^\dagger]= q^{-2N_k}.
\end{equation*}
These equations are the defining relations of the so-called $q$-boson
algebra \cite{q-bozon-algebra}. The canonical Bose
operators are recovered in the $q\to 1$ limit:
\begin{equation*}
  \lim_{q\to 1} \qB_k=a_k,\qquad
  \lim_{q\to 1} \qB_k^\dagger=a_k^\dagger.
\end{equation*}

The Hamiltonian of the $q$-boson model is defined as
\be{Hq}
H_{\text{qb}}=\sum_{j=1}^L (2N_j-\qB_j^\dagger \qB_{j+1}-\qB_{j+1}^\dagger \qB_{j}).
\ee
We assume periodic boundary conditions.

The Hamiltonian \eqref{Hq} has the form of a free
hopping model, but there are interactions between the particles due to the
fact that the $\qB$ and $\qB^\dagger$ are deformed annihilation and creation operators: their action
depends on the local occupation numbers.

\subsection{Algebraic Bethe Ansatz}

The $q$-boson model can be solved by the Algebraic Bethe Ansatz (ABA)
\cite{Faddeev:1979gh,Faddeev-ABA-intro}, which we now review. The main objects in ABA
are the monodromy matrix and its matrix elements. They are 
constructed as follows. 

Let us first consider an auxiliary space
$V_a=\complex^2$. The so-called Lax operator $\mathcal L(\lambda)$ is a linear operator acting on $V_a\otimes
V_j$, where $V_j$ is one of the local bosonic spaces. In this model the Lax operator is written in
as
\be{eredetiL}
\mathcal  L_j(\lambda)=
\begin{pmatrix}
 e^{i\lambda} & \chi \qB^\dagger_j \\
\chi \qB_j & e^{-i\lambda} 
\end{pmatrix}.
\ee
Here the matrix structure corresponds to the auxiliary space and the matrix elements are operators
acting on $V_j$. The numerical parameter $\chi$ is given by $\chi^2=1-q^{-2}$.

The Lax operator satisfies the RLL exchange relation
\be{RLL}
R(\lambda-\mu) \big(\mathcal  L(\lambda)\otimes \mathcal L(\mu)\big)=
\big(\mathcal L(\mu)\otimes \mathcal  L(\lambda)\big)  R(\lambda-\mu)
\ee
with the $R$-matrix
\be{R}
R(u-v)=
\begin{pmatrix}
f(u,v) & 0 & 0 & 0 \\
0 &  q& g(u,v) & 0 \\
0 &  g(u,v)& q^{-1}&  0 \\
0 & 0 & 0 & f(u,v) \\
\end{pmatrix},
\ee
where
\be{f-func}
{  f(u,v)=\frac{\sin(u-v+i\eta)}{\sin(u-v)},\qquad  g(u,v)=\frac{\sin(i\eta)}{\sin(u-v)}.}
\ee
This representation of the $R$-matrix is slightly different from the conventional one, due to
the factors of $q$ and $q^{-1}$. Nevertheless it satisfies the usual
Yang-Baxter relation
\begin{multline}
  \label{YB2}
    R_{1,2}(\lambda_{1,2})R_{1,3}(\lambda_{1,3})R_{2,3}(\lambda_{2,3})\\
=  R_{2,3}(\lambda_{2,3}) R_{1,3}(\lambda_{1,3}) R_{1,2}(\lambda_{1,2}),
\end{multline}
where it is understood that $\lambda_{j,k}=\lambda_j-\lambda_k$. The relation between this
$R$-matrix and its conventional form is discussed in Appendix \ref{sec:similarity}.

The monodromy matrix is constructed as
\be{eq:T}
 T(\lambda)
=\mathcal L_L(\lambda)\mathcal L_{L-1}(\lambda)\dots \mathcal L_1(\lambda) =
 \begin{pmatrix}
   A(\lambda) & B(\lambda) \\
C(\lambda) & D(\lambda) 
 \end{pmatrix}.
\ee
It follows from \eqref{RLL} that the monodromy matrix satisfies the RTT-relation
\be{RTT}
 R(\lambda-\mu) \big(T(\lambda)\otimes T(\mu)\big)=
\big(T(\mu)\otimes T(\lambda)\big)  R(\lambda-\mu).
\ee
A direct consequence of \eqref{RTT} is that the transfer matrices
defined as
\begin{equation*}
  t(\lambda)=\text{Tr } T(\lambda)=A(\lambda)+D(\lambda)
\end{equation*}
form a commuting family:
\be{comm}
  [t(\lambda),t(\mu)]=0.
\ee
The asymptotic behavior at $\lambda=\pm i\infty$ is given by
\be{vegtelen}
\lim_{\lambda\to i\infty} \Big(e^{iL\lambda}t(\lambda)\Big)=
 \lim_{\lambda\to -i\infty} \Big(e^{-iL\lambda}t(\lambda)\Big)=1.
\ee

In this model the commuting set of charges are obtained from the expansion of $t(\lambda)$ around the
special points $\lambda=i\infty$; the natural expansion parameter is $e^{2i\lambda}$. 
We define a set of charges $I_\alpha$ with $m\ge 1$ as
\be{Idef}
 I_\alpha=\frac{\alpha}{(2\alpha)! (1-q^{-2\alpha})}\left(\frac{\partial}{\partial x}\right)^{2\alpha}
\left.\log\left(x^L t(\log(x))\right)\right|_{x=0},
\ee
where $x=e^{i\lambda}$. The pre-factors are chosen such that the eigenvalues of the charges will take
a simple form.

It can be seen from the definition of the transfer matrix, that each operator  $I_\alpha$ is extensive
and it is a sum of local 
operators which span at most $\alpha+1$ sites. In particular for the first charge we obtain
\be{charges}
I_1=\sum_j \qB_j^\dagger \qB_{j+1}.
\ee
Explicit formulas for $I_2$ and $I_3$ (in a slightly different multiplicative normalization) can be
found in \cite{zvonarev-g3}. 

The operators $I_\alpha$ are not Hermitian. We can define the charges with negative indices as their adjoint:
\be{I-1}
  I_{-\alpha}=(I_\alpha)^\dagger.
\ee
They can be obtained by expanding the transfer matrix around
$\lambda=-i\infty$. Hermitian charges are then obtained as the
combinations
\begin{equation}
  \frac{I_\alpha+I_{-\alpha}}{2},\qquad  \frac{I_\alpha-I_{-\alpha}}{2i}.
\end{equation}

The particle number operator
\begin{equation*}
  N=\sum_j N_j
\end{equation*}
commutes with all of the charges, because the transfer matrix only includes terms with an equal number of
$\qB^\dagger$ and $\qB$ operators. We define $I_0\equiv N$. The Hamiltonian can then be written as
\be{energy}
  H_{\text{qb}}=-I_1-I_{-1}+2I_0.
\ee

We build the following set of vectors (called {\it Bethe vectors}) via the $B$-operators of the
monodromy matrix:
\be{Bstate}
  \ket{\{\lambda\}}\equiv \prod_{j=1}^N B(\lambda_j) \ket{0}.
\ee
Here $\ket{0}=\otimes_{j=1}^L \ket{0}_j$ is the vacuum state. The parameters $\lambda_j$ are the
rapidities of the interacting bosons.

The dual vectors can be defined as
\be{Cstate}
\bra{\{\lambda\}}\equiv  \bra{0} \prod_{j=1}^N C(\lambda_j),\qquad \bra{0}=\ket{0}^{\dagger}.
\ee

A state of the form
\eqref{Bstate} is an eigenstate of the transfer matrix if the
rapidities satisfy the Bethe equations:
\be{Be0}
{  e^{2i\lambda_j L}}\prod_{k\ne j} 
\frac{\sin(\lambda_j-\lambda_k-i\eta)}{\sin(\lambda_j-\lambda_k+i\eta)}=1.
\ee
All solutions to this equation consist of purely real rapidities.

If the Bethe equations are
satisfied, then we call the states \eqref{Bstate}-\eqref{Cstate} on-shell; in other cases we refer to them as
off-shell Bethe states.

The eigenvalues $\Lambda(u|\{\lambda\})$ of the transfer matrix $t(u)$ on the Bethe states are
\begin{multline}
  \label{tau}
  \Lambda\left(u|\{\lambda\}\right)\\=
 \frac{1}{q^N}\left(
   e^{iLu} \prod_{j=1}^N f(u,\lambda_j)+e^{-iLu}
   \prod_{j=1}^N f(\lambda_j,u)
 \right).
\end{multline}
Eigenvalues of the local charges are easily obtained using the
definition \eqref{Idef}. 
It is easy to see that they can be expressed as sums of single
particle eigenfunctions:
\be{hahaha}
  I_\alpha\ket{\{\lambda\}} =
\sum_{j=1}^N i_\alpha(\lambda_j){|\{\lambda\}\rangle},
\ee
where
\begin{equation*}
  i_\alpha(\lambda) =\frac{\alpha}{(2\alpha)! (1-q^{-2\alpha})} \left(\frac{\partial}{\partial \xi}\right)^{2\alpha}
\left.\log\left(f(\lambda,\log(\xi))\right)\right|_{\xi=0}.
\end{equation*}
In \eqref{hahaha} we used that the charge $I_\alpha$ only exists in
lattices with $L>m$, therefore it is enough to keep the second term
from \eqref{tau}. Using the substitution $e^{i\lambda}=a$ the derivatives are
calculated easily and we find
\begin{equation*}
\begin{split}
   i_\alpha(\lambda) 
= e^{-2i\alpha\lambda}.
\end{split}
\end{equation*}
It follows from \eqref{energy} that the one-particle energy  is
\be{elambda}
  e(\lambda)=2(1-\cos(2\lambda)),
\ee
which is always non-negative for the physical rapidities when $\lambda$ is purely real.

\subsection{Integrability breaking}

We wish to compute the lattice regularization of relation \eqref{Xderiv3}, which describes the
atomic losses of the Lieb-Liniger model. We choose the Lindblad jump operators
\be{qLindblad}
  L_j=\qB_j^K,\qquad  L^\dagger_j=\qB^{\dagger K}_j,
\ee
where now $j$ is a site index.

The time derivative of the charge densities is then given by
\be{Xderiv4}
\frac{d}{dt} \frac{\vev{I_\alpha}}{L}= G
\vev{\qB^{\dagger K}_1  \left[I_\alpha,\qB_1^K\right]}.
\ee
Here we assumed a homogeneous situation and chose $j=1$ for the site index.

The expectation value above concerns any equilibrium state.
We approach it from a finite volume situation, and we intend to
compute the normalized amplitude
\be{Oadef}
 \mathfrak O_\alpha=   \frac{\bra{\{ \lambda\}}\qB^{\dagger K}_1 [I_\alpha, \qB_1^K] \ket{\{ \lambda\}}}
    {\skalarszorzat{\{\lambda\}}{\{\lambda\}}},
  \ee
where the states are given by \eqref{Bstate}-\eqref{Cstate}.
  
The simplest case is that of $I_0=N$, for which we find
\begin{equation}
  \label{O0k}
   \mathfrak O_0=-Kg_{K,\text{qb}}
 \end{equation}
where we introduced the local correlation function of the $q$-boson model as
\begin{equation}
  \label{gkqb}
  g_{K,\text{qb}}=\frac{ \bra{\{ \lambda\}}\qB^{\dagger K}_1  \qB_1^K \ket{\{\lambda\}}}
    {\skalarszorzat{\{\lambda\}}{\{\lambda\}}}.
\end{equation}
In deriving \eqref{O0k} we used
\be{NKQB}
  [N, \qB_1^K]=-K \qB_1^K.
\ee

The next simplest Hermitian charges are the lattice momentum and the energy. We focus on space
reflection symmetric configurations, thus we are only interested in the energy. It follows from
\eqref{energy} that
\begin{equation}
  \frac{\bra{\{\lambda\}}\qB^{\dagger K}_1 [H_{\text{qb}}, \qB_1^K] \ket{\{\lambda\}}}
    {\skalarszorzat{\{\lambda\}}{\{\lambda\}}}=
   - 2K g_K - \mathfrak O_1- \mathfrak O_1^*.
\end{equation}
The star denotes complex conjugation.

In the next Section we will show that $g_K$ can be computed within the ABA; the concrete steps of
the present derivation are different from those of \cite{sajat-XXZ-to-LL}. Afterwards, the next step is
the computation of $\mathfrak O_1$, which requires the treatment of $I_1$ within ABA.
  As explained in the previous Section, the quantities $\mathfrak O_\alpha$ would be relatively easy
  to compute for arbitrary $\alpha$, if there were an asymptotic
  inverse for the transfer matrix.
However, to our
best knowledge there is no asymptotic inverse in the $q$-boson model, and we need an alternative approach.

Our solution is to expand the transfer matrix itself to get the charges. Using the asymptotic
behavior \eqref{vegtelen} we find the first non-trivial term as
\be{I1}
 I_1= \chi^{-2}   \lim_{u\to i\infty} e^{-2iu} \left[ e^{iLu} t(u)-1\right].
\ee
Thus the charge $I_1$ and therefore also the Hamiltonian can be treated with a single insertion of a
transfer matrix into the commutator \eqref{Oadef}. 

Higher charges could be obtained by further expanding $t(u)$ into powers of $e^{2iu}$,
and subtracting the contributions from lower order terms. For example the expansion of $t(u)$ at
order $e^{4iu}$ includes terms proportional to $I_2$ and also $(I_1)^2$. The latter term can be
subtracted using \eqref{I1}. This way the resulting expression for $I_2$ would only involve a product of
at most two transfer matrices.

In this work we content ourselves with the treatment $I_1$, and we give some further comments about
the higher charges in the Discussion.

\subsection{Scaling to the Lieb-Liniger model}

\label{sec:scaling1}

The $q$-boson model can serve as a lattice
regularization of the Lieb-Liniger model. Its scaling limit was already studied in a number of
works, for example \cite{zvonarev-g3,marci-ll-quench1}. Here we
summarize the scaling procedure, 
including the scaling of the integrability breaking amplitudes.

The idea of the scaling is to perform the continuum limit and the $q\to 1$ limit simultaneously (a
simple $q\to 1$ limit would result in free bosons on the lattice).  
To this order let us choose a small parameter $\eps$. The $q$-boson model with a coupling constant $q=e^\eta$
and a finite volume $L$ 
is scaled to the 
Lieb-Liger model with coupling $c$ in a finite volume $l$, if the parameters are connected via
\be{LL_size}
  L=l/\eps,\qquad \eta=\frac{c}{2} \eps,
\ee
and then the $\eps\to 0$ limit is taken. In this procedure the parameters $l$ and $c$ of the
Lieb-Liniger model are kept finite, the length of the lattice is increased (continuum limit) and the
coupling is scaled to zero.
During this process the rapidities $\lambda$ of the $q$-boson model will be scaled as
\be{LL_rapidities}
  \lambda={\frac{p\eps}{2}}, 
\ee
where $p$ is the rapidity parameter of the Lieb-Liniger model. Furthermore, the Lieb-Liniger space
coordinate $x$ is obtained from the site index $j$ as
\be{THE_scaling}
  x=\eps j.
\ee
The scaling of the constant $\chi$ is given by
\be{chi_scale}
\chi^2=1-e^{-2\eta}\to c\eps.
\ee

It can be seen that this scaling works on the level of wave functions and even in ABA. For example, the
scaling limit of the Bethe equations \eqref{Be0} becomes
\be{BeLL}
e^{ip_jl} \prod_{k\ne j}\frac{p_j-p_k-ic}{p_j-p_k+ic} =1.
\ee
The one-particle energy \eqref{elambda} is scaled as
\be{energy_scaling}
 e(\lambda)\quad \longrightarrow \quad\eps^2 e(p),
\ee
where $e(p)$ is the single particle energy of the Lieb-Liniger model given by
\be{energyLL}
  e(p)=p^2.
\ee

Regarding the $K$-body annihilation and creation processes we have
\be{qb1}
  \qB_1^{\dagger K} \qB_1^K\quad\longrightarrow\quad  \eps^K \Psi^{\dagger K}(0)\Psi^K(0).
\ee
This follows simply from the normalization of the field operators. As an effect, the scaling of the
local correlator becomes
\be{gkscaling}
  g_{K,\text{qb}}\quad\longrightarrow\quad   \eps^K g_{K}.
\ee

Finally the scaling of the Lindblad amplitude for the energy becomes
\begin{multline}
  \label{Hlossscaling}
    \frac{\bra{\{\lambda\}}\qB^{\dagger K}_1 [H_{\text{qb}}, \qB^K_1] \ket{\{\lambda\}}}
    {\skalarszorzat{\{\lambda\}}{\{\lambda\}}} \quad\longrightarrow
    \quad\\ 
\longrightarrow\quad
\eps^{K+2}\frac{\bra{\{p\}}\Psi^{\dagger K}(0) [H_{\text{LL}}, \Psi^K(0)] \ket{\{p\}}}
    {\skalarszorzat{\{ p\}}{\{p\}}}.
\end{multline}
Our goal is to compute the l.h.s. above exactly, in a finite volume and with a finite number of
particles. Afterwards we perform the scaling limit, and extract the leading terms, which are
of the order $\eps^{K+2}$. We take the thermodynamic limit only at the end of the computation, when the finite volume
result in the Lieb-Liniger model is already obtained.

\section{Particle losses in the $q$-boson model}

\label{sec:qloss}

In this Section we compute the loss amplitudes for the first two charges $I_0$ and $I_1$ of the
$q$-boson model. 

\subsection{Notations}

\label{sec:notations}

The computations to be presented are rather technical, with the formulas
becoming quite lengthy. In order to shorten the formulas we introduce the following special notations.

Sets and ordered sets are to be denoted as $\bx=\{x_1,\dots,x_N\}$. Omission of elements is denoted as
$\bx_k=\bx\setminus x_k$, $\bx_{k,\ell}=\bx\setminus \{x_k,x_{\ell}\}$.

For arbitrary functions $G(x)$, $F(x,y)$ and arbitrary sets $\bx$, $\by$ we define
\be{products}
G(\bx)=\prod_{i=1}^{N}G(x_i),\qquad F(\bx,y)=\prod_{i=1}^{N}F(x_i,y),\qquad \mbox{etc.}
\ee
Similarly, for products with omissions we define
\be{productso}
G(\bx_k)=\mathop{\prod_{i=1}^{N}}_{i\ne k}G(x_i),\qquad
F(\bx_k,y)=\mathop{\prod_{i=1}^{N}}_{i\ne k}F(x_i,y),\qquad \mbox{etc.}
\ee
Furthermore, for any function $g(x,y)$ of two variables we define
\begin{equation}
  \Delta_g(\bar x)=\prod_{k>j} g(x_k,x_j),\qquad
    \Delta'_g(\bar x)=\prod_{k<j} g(x_k,x_j).
\end{equation}

Now we give a few simple examples for the use of these notations. For example, in these notations
the Bethe vectors \eqref{Bstate} are
defined as
\be{Bstate2}
  \ket{\{\lambda\}}\equiv B(\bar\lambda) \ket{0}.
\ee

The Bethe equations
\eqref{Be0} can be written as
\be{Bethe_EQ}
{e^{-2i\lambda_j L} }\frac{f(\lambda_j,\bar\lambda)}{f(\bar\lambda,\lambda_j)}=-1.
\ee
Furthermore,  the transfer matrix eigenvalues \eqref{tau} are expressed as
\be{tau-short}
 \Lambda(u|\{\lambda\})=
\frac{1}{q^N}\left(
e^{iLu} f(u,\bar\lambda)+e^{-iLu}  f(\bar\lambda,u)
\right).
\ee

The strength of the bar-notation becomes evident as the formulas become more and more complicated.

\subsection{Action of $\qB$ and $\qB^{\dagger}$}

In order to compute the loss amplitudes the first step is to derive of the action of multiple $\qB$
operators on the Bethe states. This is a straightforward procedure, but for this particular model it
has not yet been done in the literature. Therefore we detail some of the steps.

From the explicit form of Lax operator \eqref{eredetiL} it is easy to establish
\be{BCfields}
\begin{split}
\chi   \psi_1&= {\lim_{v\to -i\infty}  e^{-iv(L-1)} C(v),}\\
\chi   \psi_1^\dagger&={ \lim_{v\to i\infty}  e^{iv  (L-1)} B(v)}.
\end{split}
\ee
Thus we first need the action of $C$ operators on the Bethe states.

From the RTT-relation \eqref{RTT} the following commutation relation can be found
\begin{multline}
\label{CD}
  C(\lambda)B(\mu)-e^{-2\eta}B(\mu) C(\lambda)\\
  =e^{-\eta} g(\lambda,\mu){(A(\lambda)D(\mu)-A(\mu)D(\lambda))}.
\end{multline}
Furthermore the exchange of the $A$ and $D$ operators with the $B$ operators is given by
\be{ABDB}
\begin{split}
e^\eta A(\lambda)B(\mu)&=f(\lambda,\mu)B(\mu)A(\lambda)+g(\mu,\lambda)B(\lambda)A(\mu),\\
e^\eta D(\lambda)B(\mu)&=f(\mu,\lambda)B(\mu)D(\lambda)+g(\lambda,\mu)B(\lambda)D(\mu).
\end{split}
\ee

Using \eqref{CD} the formula of action of $\qB$ operators on the Bethe vectors can be found
explicitly. The computation is presented in Appendix \ref{homogeneous2}. The result is
\begin{multline}
\label{CC_red}
\psi_1^K|\{\mu\}\rangle={\frac{\chi^{K} e^{\eta \left(K(K+1)/2-N  K\right)}}{(2\sinh(\eta))^K}}
\prod_{\ell=1}^K\left(1-e^{-2\ell\eta}\right) \\
\times \sum {a(\bmu_{\so})e^{i\bmu_{\so}} }f(\bmu_{\so},\bmu_{\st})|\{\mu_{\st}\}\rangle.
\end{multline}
The summation here is taken over all possible partitions $\bar\mu\to\{\bar\mu_{\so},\bar\mu_{\st}\}$
such  that $|\bmu_{\so}|=K$. We expanded our agreement for the shorthand notation as
\be{eprod}
e^{\bmu}=\prod_{\mu_j\in \bmu}e^{\mu_j}.
\ee
Furthermore we will use the shorthand notation for the prefactor
\be{CK}
C_K=\frac{1}{\left(2\sinh(\eta)\right)^K}\prod_{\ell=1}^K\left(1-e^{-2\ell\eta}\right).
\ee

\subsection{Norm}

The norm of an eigenvector $\ket{\{\mu\}}$ of the $q$-boson model is given by \cite{Korepin-Book}
\be{norm}
\langle\{\mu\}|\{\mu\}\rangle=\Delta_g(\bmu)\Delta_g'(\bmu)h(\bmu,\bmu){\sinh^N(\eta)}\det \mG\left(\{\mu\}\right),
\ee
with
\be{gh}
h(\mu,\lambda)=f(\mu,\lambda)g^{-1}(\mu,\lambda).
\ee
The so-called Gaudin matrix is defined as
\begin{multline}
\label{Gaudin}
\mG_{jk}(\{\mu\})=\delta_{jk}\left(L+\sum_{\ell=1}^a\varphi_q(\mu_k-\mu_{\ell})\right)-\varphi_q(\mu_k-\mu_j),\\
j,k=1,\dots,N,
\end{multline}
with
\be{phi_scattering}
\varphi_q(\mu)={\frac{\sinh(2\eta)}{\sin(\mu+i\eta)\sin(\mu-i\eta)}}.
\ee

\subsection{Scalar products}  

As a next step we compute normalized scalar products of the type
\be{scalp_norm}
\frac{\bra{\{\mu\}}\qB_1^{\dagger K}|\{\mu_{\st}\}\rangle}{\skalarszorzat{\{\mu\}}{\{\mu\}}},
\end{equation}
where $\bar\mu_{\st}$ is a subset of $\bar \mu$ with $|\bar \mu_{\st}|=N-K$. In the actual
computation of the loss amplitudes we will need such scalar products with the number of field
operators given by $K$ and $K+1$.

The normalized scalar products above can be
found from the definition \eqref{BCfields} by using the famous Slavnov formula for the
overlaps between on-shell and off-shell Bethe states. The detailed computation is given in Appendix
\ref{homogeneous1}. The result is
\begin{multline}
\label{scalarproduct}
\frac{\bra{\{\mu\}}\qB_1^{\dagger K}|\{\mu_{\st}\}\rangle}{\skalarszorzat{\{\mu\}}{\{\mu\}}}=
\prod_{j=1}^K \left(1-e^{-2j\eta}\right) { \frac{ \chi^{-K} e^{KN\eta}}{(2\sinh(\eta))^{K(K-1)/2}} }\\
\times  \frac{ {e^{iK\bmu_{\so}}}  }
{ d(\bar\mu_{\so})   \Delta_g(\bar\mu_{\so}) f(\bar\mu_{\st},\bar\mu_{\so})    h(\bar\mu_{\so},\bar\mu_{\so}) }
\frac{\det {\mathcal{\tilde M}\left(\{\mu_{\st}\}|\{\mu\}\right)}}{\det \mathcal{G}(\{\mu\})}.
\end{multline}
Here $\mathcal {\tilde M}$ is a matrix obtained after a slight modification of $\mG$ (see \eqref{Mtilde}). 

\subsection{Local correlation function}

Let us first compute the quantity $g_{K,\text{qb}}$ defined in \eqref{gkqb}. 
To do this we take the scalar product between $\bra{\{\mu\}}\qB_1^{\dagger K}$ and the r.h.s. of
\eqref{CC_red}. This gives
\begin{multline}
\label{gkqb2}
g_{K,\text{qb}}=
{C_K\chi^{K}}e^{\eta \left(K(K+1)/2-N K\right)} \\\
\times{e^{-\eta K}} \sum {a(\bmu_{\so})e^{i\bmu_{\so}}}f(\bmu_{\so},\bmu_{\st})
\frac{\bra{\{\mu\}}\qB_1^{\dagger K}|\{\mu_{\st}\}\rangle}{\skalarszorzat{\{\mu\}}{\{\mu\}}}.
\end{multline}
Summation is taken over partitions  $\bmu\to\{\bmu_{\so},\bmu_{\st}\}$, such that
$|\bmu_{\so}|=K$.

Substituting \eqref{scalarproduct} into \eqref{gkqb2} we obtain after elementary simplifications
\begin{multline}
\label{gkqb1b}
g_{K,\text{qb}}=  {\prod_{j=1}^K \left(1-e^{-2j\eta}\right)}\frac{C_K 2^K}{\chi^{K(K+1)}} \times \\
\times
\sum s\left(\{\mu_{\so}\}\right)
\frac{\det \mG^{(f)}}{\det \mG},
\end{multline}
where
\begin{multline}
\label{Gf}
\mG^{(f)}_{jk}={\sinh(\eta)e^{-i(2k-K-1)\mu_j} },\qquad k=1,\dots,K,\\
\mG^{(f)}_{jk}=\mG_{jk},\qquad k=K+1,\dots,N,\phantom{11111111111.}
\end{multline}
and $s(\{\mu\})$ is a function defined as
\begin{multline}
\label{S-def}
s\left(\{\mu\}\right)=\frac{1}{\Delta_g(\bmu)\Delta_h(\bmu)\Delta_h'(\bmu)}=\\
=\prod_{j<k}\frac{\sin(\mu_{jk}){\sin(i\eta)}}{\sin(\mu_{jk}+i\eta)\sin(\mu_{jk}-i\eta)}.
\end{multline}
The elements of the matrix  $\mG^{(f)}$ are complex, but the phases are arranged symmetrically, and it can be seen that the answer is manifestly real.

The result \eqref{gkqb1b} is analogous to similar formulas found in \cite{sajat-XXZ-to-LL}. Regarding
the $q$-boson model it is new.

\subsection{Action of transfer matrix}

In order to compute the loss amplitudes for the conserved charges we also need the action of
TM's on off-shell Bethe states. For the charge $I_1$ (and thus the energy) it is enough
to take a single TM, which acts as
\be{t_action}
t(u)\ket{\{\mu\}}=\Lambda\left(u|\{\mu\}\right)\ket{\{\mu\}}+
\sum_j\tilde\Lambda_j\left(u|\{\mu\}\right)
\ket{\{\mu_j,u\}}.
\ee
Here the first term on the r.h.s. is the so-called ``wanted term'' which gives the TM eigenvalue $\Lambda(u|\bar\mu)$
if the Bethe
vector is on-shell.
The remaining terms are the ``unwanted terms'' for which the pre-factors are
\begin{multline}
\tilde \Lambda_j\left(u|\{\mu\}\right)= e^{-\eta N}\Big[  a(\mu_j) g(\mu_j,u)f(\mu_j,\bar \mu_j)\\
+d(\mu_j) g(u,\mu_j)f(\bar \mu_j,\mu_j)\Big].
\end{multline}

Now plugging \eqref{CC_red}, \eqref{scalarproduct} and  \eqref{t_action}   into \eqref{Oadef} and \eqref{I1} we can write 
\begin{widetext}
\begin{multline}
\label{I_loss}
\mathfrak O_1=\lim_{\omega\to i\infty}e^{i(L-2)\omega}\left\{\frac{\bra{\{
    \mu\}}\psi^{\dagger K}_1 t(\omega) \psi_1^K\ket{\{\mu\}}}{\langle\{\mu\}|\{\mu\}\rangle}-
 \frac{ \bra{\{\mu\}}\psi^{\dagger K}_1 \psi_1^K \ket{\{\mu\}} }{\langle\{\mu\}|\{\mu\}\rangle}  \Lambda\left(\omega|\{\mu\}\right)\right\}\\
=C_K\chi^K e^{-(N-(K+1)/2)K\eta}\sum
 a(\bar \mu_{\so})e^{i\bar \mu_{\so}}  f(\bar\mu_{\so},\bar \mu_{\st}) \\
\times \lim_{\omega\to i\infty}e^{i(L-2)\omega}\left[\frac{
 \bra{\{\mu\}}\psi^{\dagger K}_1 \ket{\{\mu_{\st}\}}}{{\langle\{\mu\}|\{\mu\}\rangle}}  \left(\Lambda(\omega|\{\mu_{\st}\})-  \Lambda(\omega|\{\mu\})\right)
+\sum_j \tilde\Lambda_j(\omega|\{\mu_{\st}\})\frac{\bra{\{\mu\}}\psi^{\dagger K}_1 \ket{\{\bmu_{\st,j},\omega\}}}{{\langle\{\mu\}|\{\mu\}\rangle}}  
 \right].
\end{multline}
\end{widetext} 
Here the outer summation is taken over partitions $\bmu\to\{\bmu_{\so},\bmu_{\st}\}$ with
$|\bmu_{\so}|=K$, and the inner summation is over the partitions
$\bmu_{\st}\to\{\bmu_{\st,j},\mu_j\}$ where $|\mu_j|=1$ and $\bmu_{\st,j}=\bmu_{\st}\setminus\mu_j$. 

The limit $w\to i\infty$ in the direct amplitude follows from \eqref{I1}. The limit $w\to i\infty$
of the indirect amplitude can be computed using
\begin{multline}
\lim_{u\to i\infty} e^{ iu} \tilde \Lambda_j(u|\bar \mu_{\st})=2e^{-(N-K)\eta}\sinh(\eta)e^{i\mu_j}\\
\times \left[  a(\mu_j) \frac{f(\mu_j,\bar \mu_j)}{f(\mu_j,\bar \mu_{\so})}-
d(\mu_j) \frac{f(\bar \mu_j,\mu_j)}{f(\bar\mu_{\so},\mu_j)}\right].
\end{multline}

\subsection{Loss amplitude of $I_1$}

We can now rewrite \eqref{I_loss} as the sum of two parts
\begin{equation}
\mathfrak O_1=\dire+\indire,
\end{equation}
where  we call $\dire$ and $\indire$ the {\it direct term} and {\it indirect term},
respectively. The direct term stems from the direct action of the transfer matrix on the eigenstate,
and it is given by 
\begin{multline}
\label{Direct}
\dire=-{C_K e^{-(N-(K+1)/2)K\eta} }\\
 \times\sum a(\bar \mu_{\so})e^{{i}\bar \mu_{\so}}  f(\bar\mu_{\so},\bar \mu_{\st})
 \frac{ \bra{\{\mu\}}\psi^{\dagger K}_1 \ket{\{\mu_{\st}\}}}{\skalarszorzat{\{\mu\}}{\{\mu\}}}
 I_1\left(\{\mu_{\so}\}\right).
\end{multline}
Here the summation is taken over partitions  $\bmu\to\{\bmu_{\so},\bmu_{\st}\}$, such that
$|\bmu_{\so}|=K$, and for the eigenvalues of $I_1$ we use the following shorthand notation 
\be{I_def}
I_1\left(\{\mu_{\so}\}\right)=\sum_{\mu_j\in\bmu_{\so}}e^{2{ i}\mu_j}.
\ee

The indirect term results from the ``unwanted terms'' of the action of the transfer matrix, and it reads
\begin{multline}
\label{Indirect}
\indire= {C_K} e^{-(N-(K+1)/2)K\eta} \chi e^{-(N-K-1)\eta} \\
\times  \sum
  a(\bar \mu_{\so})a(\mu_b) e^{i\bar \mu_{\so}}  e^{i\mu_b}
  f(\bar\mu_{\so},\bar \mu_{\st})f(\mu_b,\bar \mu_{\st}) \\
\times  \left[  f(\bar\mu_{\so},\mu_b)- f(\mu_b,\bar\mu_{\so})\right]
  \frac{\bra{\{\mu\}}\psi^{\dagger (K+1)}_1 \ket{\{\mu_{\st}\}}}{\skalarszorzat{\{\mu\}}{\{\mu\}}}.
\end{multline}
The summation is taken over all partitions  $\bmu\to\{\bmu_{\so},\bmu_{\st},\bmu_b\}$, such that
$|\bmu_{\so}|=K$, $\mu_b$ is a single rapidity, and finally $|\bmu_{\st}|=N-K-1$. Note that the set
$\bmu_{\st}$ here is different than in the previous formula: now it has one less
element.

To proceed we substitute the scalar products \eqref{scalarproduct} with $K$ and $K+1$ into the
formulas above.

For the direct term \eqref{Direct} we obtain after elementary simplifications 
\begin{multline}
\label{Y1}
\dire=-C_K\prod_{j=1}^K \left(1-e^{-2j\eta}\right)\frac{2^K}{(2\sinh(\eta))^K \chi^{K(K+1)}} \\
\times\sum s\left(\{\mu_{\so}\}\right)\frac{\det \mG^{(f)}}{\det \mG}I_1\left(\{\mu_{\so}\}\right).
\end{multline}
Here summation is taken over the same partitions as in \eqref{Direct}.

Substituting \eqref{B1R} into \eqref{Indirect} we get
\begin{multline}
\label{Z1}
\indire=C_K\prod_{j=1}^{K+1} \left(1-e^{-2j\eta}\right)   \frac{2^{K+1}   }{\chi^{(K+1)(K+2)}}\\
\times \sum\frac{e^{{2i}\mu_b} }{\Delta_g(\bar\mu_{\so})h(\bar\mu_{\so},\bar\mu_{\so}) 
h(\mu_b,\bar\mu_{\so})}\\
\times \left[ 1-\frac{f(\mu_b,\bar\mu_{\so})}{f(\bar\mu_{\so},\mu_b)}\right]\frac{\det \mG^{(f)}}{\det \mG},
\end{multline}
where 
\begin{equation}
\begin{split}
\mG^{(f)}_{jk}&=\sinh(\eta)e^{{-i(2k-K-2)\mu_j}},\qquad k=1,\dots,K,\\
\mG^{(f)}_{jk}&=\sinh(\eta)e^{{-i(2k-K-2)\mu_b}},\qquad k=K+1,\\
\mG^{(f)}_{jk}&=\mG_{jk},\qquad k=K+2,\dots,N.
\end{split}
\end{equation}
Here rapidities $\{\mu_k\}$ in the first $K$ columns belong to the set $\bmu_{\so}$. Summation is taken over the same partitions as in \eqref{Indirect}.

\subsection{Energy loss}

Using \eqref{energy} and \eqref{I-1} we can present the energy loss amplitude as a sum of two terms:
\begin{equation}
\frac{\bra{\{\mu\}}\psi^{\dagger K} [H_{\text{qb}}, \psi_1^K]\ket{\{\mu\}}}{\skalarszorzat{\{\mu\}}{\{\mu\}}}
= - (\dire_{E,K}+\indire_{E,K}).
\end{equation}
Here the subscript $E$ denotes that the quantities describes the energy
loss, and $K$ stands for the $K$-body processes.

The new direct term is:
\begin{multline}
 \dire_{E,K}=
  {C_K}   \prod_{j=1}^K \left(1-e^{-2j\eta}\right)
 \frac{2^K}{\chi^{K(K+1)}}\\
\times\sum  E(\{\mu_{\so}\}) s(\{\mu_{\so}\})
\frac{\det \mG^{(f)}}{\det \mG},
\end{multline}
where $E(\{\mu_{\so}\})$ is given by the sum of the one-particle eigenvalues \eqref{elambda}
\be{energy_lambda}
E(\{\mu_{\so}\})=\sum_{\mu_j\in\bmu_{\so}}^Ne(\mu_j).
\ee

The new indirect term is given by
\begin{multline}
\indire_{E,K}={C_K}
\prod_{j=1}^{K+1}\left(1-e^{-2j\eta}\right)  \frac{{2^{K+1}}}{\chi^{(K+1)(K+2)}}\\
\times \sum s(\{\mu_{\sth}\}) \frac{\det \mG^{(f)}}{\det \mG} \\
\times \left[\sum\left[f(\bmu_{\so},\mu_b)- f(\mu_b,\bmu_{\so})\right]\left(e^{{ 2i}\mu_b}-e^{-{ 2i}\mu_b}\right)\right],
\end{multline}
here we denote set $\bmu_{\sth}=\{\bmu_{\so},\mu_b\}$, $|\bar p_{\sth}|=K+1$ and use the rewriting $h(\bmu_{\so},\mu_b)h(\mu_b,\bmu_{\so})h(\bmu_{\so},\bmu_{\so})=h(\bmu_{\sth},\bmu_{\sth})$. The first sum is taken over partitions 
$\bmu\to\{\bmu_{\st},\bmu_{\sth}\}$, $|\bmu_{\sth}|=K+1$ and the second over partitions $\bmu_{\sth}\to\{\bmu_{\so},\mu_b\}$.

This way the indirect term becomes
\begin{multline}
\indire_{E,K}={C_K}
\prod_{j=1}^{K+1}\left(1-e^{-2j\eta}\right)
\frac{{2^{K+1}}}{\chi^{(K+1)(K+2)}} \\
\times \sum_{\bmu\to\{\bar\mu_{\st}, \bmu_{\sth}\} }
s(\{\mu_{\sth}\})
F\left(\{\mu_{\sth}\}\right)\frac{\det \mG^{(f)}}{\det \mG},
\end{multline}
where the sum is taken over partitions $\bmu\to\{\bmu_{\st}, \bmu_{\sth}\}$,
$|\bmu_{\sth}|=K+1$. The function $F$ is given by
\be{F_def}
F\left(\{\mu_{\sth}\}\right)=\sum \left(e^{2{ i}\mu_b}-e^{-2{ i}\mu_b}\right)
\left[ h(\bmu_{\so},\mu_b)- h(\mu_b,\bmu_{\so})\right].
\ee
The sum in \eqref{F_def} is taken over partitions $\bmu_{\sth}\to\{\bmu_{\so}, \mu_b\}$, where
$\mu_b$ is a single rapidity.

\section{Scaling limit towards the Lieb-Liniger model}

\label{sec:scaling}

Now we derive the scaling limit of the formulas of the previous Section, using the scaling rules
given in Section \ref{sec:scaling1}.

\subsection{Local correlation function}

As a first step we derive the value of the local correlation function.

The scaling of the Gaudin matrix is given by
\be{G_ratio_scale}
  \frac{\det \mG^{(f)}}{\det \mG^{LL}}\quad\to\quad 2^{-K}(i\eps)^{K(K-1)/2}\eps^K \frac{\det \mG^{(e)}}{\det \mG^{LL}},
\ee
where $\mG^{LL}$ now is the Gaudin matrix of the Lieb-Liniger model
\be{Gaudin_LL}
\mG^{LL}_{jk}=\delta_{jk} \left[l+\sum_m \varphi(p_j-p_m)\right]-\varphi(p_j-p_k)
\ee
with
\be{varphi2}
\varphi(p)=\frac{2c}{p^2+c^2},
\ee
and {$\mG^{(e)}$ coincides with $\mG^{LL}$ except the first $K$ columns that are given by}
\be{Ge_def}
\mG^{(e)}_{jk}=(p_j)^{k-1},\quad k\le K,\quad j=1,\dots,N.
\ee

Collecting all factors we obtain 
\be{gkll}
g_K
=(K!)^2 
 \sum s(\{ p_{\so}\})
 \frac{\det \mG^{(e)}}{\det \mG^{LL}}.
\ee
Here the function $s(\{p\})$ is given by 
\be{s_new_def}
  s(\{p\})=\prod_{j>k} \frac{p_j-p_k}{(p_j-p_k)^2+c^2},
\ee
and the summation runs over the partitions $\bar p\to\{\bar p_{\so},\bar p_{\st}\}$ with $|\bar p_{\so}|=K$.
During the derivation we also used $C_K\to  K!$.

The formula above is completely identical to the earlier results of \cite{sajat-XXZ-to-LL}, in particular
it agrees with eq. (5.6) of that work.

\subsection{Energy loss rate}

Let us now focus on the scaling limit of the energy loss amplitude. 
In this case we 
expect an additional factor of $\eps^2$ according to \eqref{Hlossscaling}.

We compute the amplitude as
\be{amplitude}
  \frac{\bra{\{p\}}\Psi^{\dagger K}(0) [H_{\text{LL}}, \Psi^K(0)]\ket{\{p\}}}
  {\skalarszorzat{\{p\}}{\{p\}}}
=-(\dire_{E,K}+\indire_{E,K}),
\ee
where now $\dire_{E,K}$ and $\indire_{E,K}$ are the direct and indirect terms in the Lieb-Liniger model.

Collecting all factors
we find for the direct term
\be{YE}
    \dire_{E,K}
 =(K!)^2 
\sum  E(\{ p_{\so}\})s(\{ p_{\so}\})
\frac{\det \mG^{(e)}}{\det \mG^{LL}},
\ee
where summation is taken over partitions $\bar p\to \{\bar p_{\so},\bar p_{\st}\}$, $|\bar p_{\so}|=K$ and
\begin{equation}
  E(\{ p_{\so}\})=\sum_{p_j\in \bar p_{\so}}  (p_j)^2.
\end{equation}

Regarding the indirect term we get
\be{ZE}
  \indire_{E,K}=K!(K+1)!   \sum
F(\{ p_{\sth}\})s(\{ p_{\sth}\}) \frac{\det \mG^{(e)}}{\det \mG^{LL}},
\ee
where summation in \eqref{ZE} is taken over partitions $\bar p\to\{\bar p_{\st},\bar p_{\sth}\}$, $|\bar
p_{\sth}|=K+1$. The matrix $\mG^{(e)}$ coincides with $\mG^{LL}$ except the first $K+1$ columns where
\begin{equation}
\mG^{(e)}_{jl}= (p_j)^{l-1} ,\qquad l=1,\dots, K+1,\qquad j=1\dots N,\\
\end{equation}
and $p_j\in\bar p_{\sth}$. The function $F(\bar p_{\sth})$ after scaling is given by
\be{F_def_scale}
F(\{ p_{\sth}\})=\sum_{\bar p_{\sth}\to\{\bar p_{\so}, p_b\}}   2ip_b \left( f(\bar p_{\so},p_b)- f(p_b,\bar p_{\so})\right).
\ee
It is shown in Appendix \ref{sec:F} that this function is in fact equal to the constant
$2K(K+1)c$. Then we arrive at the simplified formula
\be{ZE2}
\indire_{E,K}=2cK((K+1)!)^2   \sum
 s(\{ p_{\sth}\}) \frac{\det \mG^{(e)}}{\det \mG^{LL}}.
\ee
The summation here is taken as in \eqref{ZE}. We can see that this expression is proportional to the $K+1$-body correlator. In fact, comparing to
\eqref{gkll} we get
\be{indirerel}
  \indire_{E,K}=2cK g_{K+1}.
\end{equation}
This result has to be compared to equation \eqref{GcK}, which was obtained from the commutation relation of
the field operators. Indeed we see that the expected higher local correlation function emerges at
the end of the computation, with the same pre-factor as obtained by \eqref{GcK}.

\section{Thermodynamic limit and factorization}

\label{sec:TDL}

Here we take the thermodynamic limit of the loss amplitudes. We take
the infinite volume limit  such that the Bethe rapidities become dense in rapidity space. We
introduce the root density $\rho(p)$, such that the number of rapidities between $p$ and $p+dp$
becomes $L\rho(p)dp$ in a large volume $L$. 

We do not specify the nature of the equilibrium state; it can be the ground state  with a given
particle density, a finite temperature state, or any other excited state which is relevant for
experiments. Therefore we also introduce the hole density $\rho_h(p)$, the total density
$\rho_t(p)=\rho(p)+\rho_h(p)$ 
and the filling fraction
\be{fill-factor}
f(p)=\frac{\rho(p)}{\rho_t(p)}.
\ee
It follows from the Bethe equations that the root and hole densities satisfy the linear integral equation
\be{rho_total}
\rho_t(p)=\frac{1}{2\pi}+\int \frac{dp'}{2\pi} \varphi(p-p') \rho(p').
\ee
The filling fraction can be use to characterize the equilibrium states.
In the case of the ground state $f(p)$ is such that it is 1 within the Fermi zone
$|p|\le p_F$ and zero otherwise. In the finite temperature case we have
$f(p)=(1+e^{\varepsilon(p)})^{-1}$, where 
$\varepsilon(p)$ is the solution of the nonlinear integral equation \cite{yang-yang-tba}
\be{NLIE}
\varepsilon(p)=\frac{p^2-\nu}{T}-\int dp'\;\varphi(p-p')\log\left(1+e^{-\varepsilon(p')}\right),
\ee
where $T$ is the temperature and $\nu$ is the chemical potential.
In quantum quench problems the filling fraction can be sometimes be found using the Quench-Action
method, see for example
\cite{caux-stb-LL-BEC-quench,lorenzo-pasquale-attractive-LL,lorenzo-pasquale-attractive-LL2}. In
this work we do not specify $f(p)$, we leave it as an arbitrary function that enters the final formulas.

In order to take the thermodynamic limit of the formulas of the previous Section we apply the
methods of \cite{sajat-XXZ-to-LL} which were based on earlier results in the literature. There are
two key steps in this procedure: the summation over partitions is turned into a multiple integral,
and the ratios of determinants are expressed using certain auxiliary functions.

Let us consider a summation over partitions $\bar p=\{\bar p_{\so},\bar p_{\st}\}$, where $|\bar p_{\st}|=K$ is
fixed, and $|\bar p_{\so}|=N-K$ is taken to infinity.
We can write simply
\be{sum-int}
  \sum_{\bar p=\{\bar p_{\so},\bar p_{\st}\}}\hdots\quad\longrightarrow\quad
\frac{L^K}{K!}  \int d\bar p_{\st}\ \rho(\bar p_{\st})\hdots
\ee
The factor of $K!$ in the denominator cancels the overcounting of the different permutations of a
given $\bar p_{\st}$ in the multiple integral.

Let us now investigate the ratios of the determinants. We apply the standard trick 
\begin{equation}
  \frac{\det \mG^{(e)}}{\det \mG^{LL}}=\det\left(\left(\mG^{LL}\right)^{-1} \mG^{(e)}\right).
\ee
The matrices in the numerator are such that they are only modified in a limited number of elements,
thus after multiplication with $\mG^{-1}$ we obtain a matrix which is equal to the identity except
for the rows affected.

In the thermodynamic limit the action of the Gaudin matrix can be transformed into an integral
equation, and we obtain the result
\begin{equation}
  \det\left(\left(\mG^{LL}\right)^{-1} \mG^{(e)}\right)=\frac{1}{(2\pi L)^K \rho_t(\bar p_{\st})}
  \times \det I,
\end{equation}
where $I$ is a matrix of size $K\times K$ with elements given by
\begin{equation}
  I_{jl}=h^{(l-1)}(p_{\st,j}).
\end{equation}
Here $h^{(l)}(p)$ are functions that are defined as
\be{hhdef}
h^{(l)}(p)=p^l +\int \frac{dp'}{2\pi} \varphi(p-p') f(p') h^{(l)}(p').
\ee
Notice that for $l=0$ we have $h^{(0)}(p)=2\pi \rho_t(p)$.

Let us now write down the multiple integrals that arise from the computation. In order to further
simplify the formulas we introduce one more notation:
\be{barpint}
  \int (d\bar p) \equiv \int \frac{d\bar p\ f(\bar p)}{(2\pi)^K}=
  \int \prod_{j=1}^K \left(\frac{dp_j}{2\pi}f(p_j)\right).
\ee
In this notation the cardinality of the set $\bar p$ is suppressed, but it is always given in the text.

In the case of the local correlator we get 
\be{GK-fin}
  g_K=(K!) \int (d\bar p)\ s\left(\{ p\}\right) \det I
\ee
with $|\bar p|=K$. The pre-factor $s(\{p\})$ is completely anti-symmetric in its
variables. After an  expansion of the determinant we can use this property to write
\be{gKfinal}
  g_K=(K!)^2 \int (d\bar p)\ s\left(\{ p\}\right) \prod_{j=1}^K h^{(j-1)}(p_j).
\ee
This agrees with the final result of the work \cite{sajat-XXZ-to-LL}.

For the indirect term the relation \eqref{indirerel} was already established in a finite volume; the
relation clearly holds also in the TDL. Thus the only remaining task is to take the limit of the
direct term. We get
\begin{multline}
\label{direct_integral}
\dire_{E,K} =(K!)^2\times \\
\times \int (d\bar p)\ s\left(\{ p\}\right)E\left(\{p\}\right) \prod_{j=1}^K h^{(j-1)}(p_j).
\end{multline}

Using \eqref{Hlossrate} the energy loss rate is eventually given by
\be{enloss}
  \frac{d}{dt} \frac{H}{L}=-G(\dire_{E,K}+2cK g_{K+1}).
\ee
We have thus obtained the energy loss rate as a sum of a $K$-fold and a $K+1$-fold integral.

~

\subsection{Factorization}

\label{sec:fac}

The multiple integrals can be factorized. Factorization of the local correlator \eqref{gKfinal} was already performed in
\cite{sajat-XXZ-to-LL}. The factorization of the direct term \eqref{direct_integral} is made in
Appendix \ref{Factorization}. Here we just present the final results, and for the sake of
completeness we also repeat here the results of \cite{sajat-XXZ-to-LL}.

The idea of the factorization is to express the multiple integral as a combination of single
integrals. It turns out that the building blocks for the final formulas are the following simple integrals:
\be{nm}
\{n,m\}=\frac{1}{c^{n+m+1}} \int \frac{dp}{(2\pi)}f(p)p^n h^{(m)}(p).
\ee
The coupling constant $c$ has the same dimension as the rapidity parameter, thus the quantities
$\{n,m\}$ are dimensionless. The addition of the pre-factor before the integral is new here, it was not
 included in the corresponding formula of \cite{sajat-XXZ-to-LL}. It can be proven that the
 following symmetry relation holds \cite{sajat-XXZ-to-LL}:
 \begin{equation}
   \{n,m\}=\{m,n\}.
 \end{equation}

 Let us now first present the factorized formulas for the local correlators. In the simplest case we have
\be{g1}
   g_1=c\{0,0\}=\frac{N}{L}.
\ee
For $K=2$ we have
\be{g2}
  g_2=2c^2\left(\{0,2\}-\{1,1\}\right).
\ee
This result can be obtained simply using the Hellman-Feynman theorem
\cite{sinhG-LL1,sinhG-LL2,g3-marci-1}.

For $K=3$ the result was first found in \cite{g3-marci-1} and it reads
\begin{multline}
\label{g3}
     g_3=
  c^3  \Big(-4 \{1,3\}+3\{2,2\}+\{0,4\}
+\{0,2\}  \\ 
 -\{1,1\}+2\left(\{0,1\}^2-\{0,0\}\{1,1\}\right)\Big).
  \end{multline}

\begin{widetext}

For $K=4$ the result was obtained in \cite{sajat-XXZ-to-LL}:
\begin{multline}
\label{g4}
    g_4=
    \frac{2}{5}c^4\Big[
8  \Big(\{0, 1\}^2  -  \{0, 0\} \{1, 1\}\Big) +  
32  \Big(\{0, 1\} \{0, 3\}-  \{0, 0\} \{1, 3\}\Big) +
 24\Big(\{0, 2\} \{1, 1\}- \{0, 1\}\{1,2\}\Big) \\
+ 30  \Big(\{0, 0\} \{2, 2\}-\{0, 2\}^2\Big)+
 4 \Big(\{0, 2\} -\{1,1\}\Big) + 
   5 \Big(\{0, 4\}  - 4 \{1, 3\}+3  \{2, 2\}\Big)\\
+  \{0, 6\}   - 6 \{1, 5\} + 15 \{2, 4\}  -  10 \{3, 3\} 
\Big].
  \end{multline}

  Let us now turn to the factorization of the direct terms. The computation is presented in Appendix
  \ref{Factorization}, the results are as follows.

  For $K=1$ we have a simple integral, which gives
\be{dE1}
  \dire_{E,1}=c^3\{0,2\}.
\ee

In case $K=2$ the answer is given by 
\be{Yfin2}
\dire_{E,2}=\frac{2c^4}{3}\left(2(\{0,4\}-\{1,3\} +\{0,0\}\{1,1\}-\{0,1\}^2)+\{1,1\}-\{0,2\}\right).
\ee
For $K=3$ we find
  \begin{equation}
    \begin{split}
      \label{Yfin}
\dire_{E,3}=\frac{c^5}{10}\Big[
  28 (\{0, 0\} \{1, 1\}- \{0, 1\}^2)
+ 84 (\{0, 1\} \{1, 2\} -  \{0, 2\} \{1, 1\}) 
+ 14 (\{1, 1\}   -  \{0, 2\}) - 5 \{0, 4\} + 9 \{0, 6\} \\
+90 (\{0, 2\}^2 - \{0, 0\} \{2, 2\})
+ 72 (\{0, 0\} \{1, 3\}- \{0, 1\} \{0, 3\})
+ 50 \{1, 3\}
  - 24 \{1, 5\} - 45 \{2, 2\}  + 
 15 \{2, 4\}
\Big].
    \end{split}
  \end{equation}
  \end{widetext}

The total rate of the energy loss is given by formula \eqref{enloss}, where the above results for
$g_K$ and $\dire_{E,K}$ have to be substituted.

{It is important that in deriving the factorized formulas we assumed that the integrals in the
definition \eqref{nm} are well defined. This is certainly true for the ground states and the finite
temperature states. However, in certain non-equilibrium situations the root distribution function
can acquire algebraic tails, making the single integrals ill defined. In these cases an appropriate
cut-off procedure is needed; we leave this question to future work.}

\subsection{Numerical evaluation}

In this paper we contain ourselves with the derivation of the exact results, and we leave the
numerical analysis of concrete cases to a future work. Nevertheless let us give some comments about
the numerical methods.

Our factorized results are expressed using the building blocks \eqref{nm} which use the auxiliary
functions \eqref{hhdef}. These are the same objects that appeared in the previous work
\cite{sajat-XXZ-to-LL}, where it was already demonstrated that they can be computed very
efficiently. The auxiliary functions are found by solving a simple linear equation, and the objects
\eqref{nm} are simple integrals over them. Therefore the quantities $\{n,m\}$ can be computed with
arbitrary numerical precision. The input to this numerical procedure is the filling fraction $f(p)$,
which enters the linear equations \eqref{hhdef}. The steps of this numerical procedure are
  standard by now, and publicly available codes can also be used as an aid \cite{ifluid}.

\section{Discussion}

\label{sec:disc}

In this work we treated the atomic losses in the repulsive Lieb-Liniger model. We explained that the
losses of the canonical charges can be computed via the $q$-boson model, which serves as a lattice
regularization of the problem. In the concrete computation we considered the energy loss, which we
expressed as a sum of two multiple integrals. The final formula is
\eqref{enloss}, which uses \eqref{gKfinal}
and \eqref{direct_integral}. The multiple integrals are
factorized explicitly for $K=1,2,3$, the formulas are presented in Section \ref{sec:fac} above.

Our work leaves a number of open questions:
Is it possible to compute
explicit formulas for the losses of the higher moments? What types of integral formulas can we
expect? Is it always possible to factorize the multiple integrals? Is there a deeper algebraic structure
behind the factorization? Can we expect to find the time derivative of the full root density, perhaps
in some approximative scheme such as a large coupling expansion?
At present we do not know the answer to these questions, but we wish to give some comments about them.

It is clear from the structure of our computation, that in principle all conserved charges can be
treated in the $q$-boson model. Even though we do not have an asymptotic inverse in this model, the
higher charges can be expressed from products of transfer matrices, thus they can be treated within
ABA. We expect multiple integral formulas with an increasing number of integrals as we move to
higher and higher charges.
It is not clear whether there should be any pattern in the resulting formulas: it is possible that
the higher charges need to be treated on a case by case basis, making the approach
impractical. Nevertheless we believe that the next charge $I_2$ could be treated with
reasonable effort, and this would yield the fourth moment of the root distribution in the
Lieb-Liniger model. 

Regarding factorization it is quite likely that all expected multiple integrals can eventually be
factorized. However, at present there is no algebraic theory for this procedure. Based on the results
of this work and \cite{sajat-XXZ-to-LL}, and also 
on experience with the Heisenberg chain it is very likely that factorization needs to be performed
on a case by case basis, and there will not be any closed formulas applicable to all the higher moments.

Summarizing these expectations, the search for the time derivative of the full root density is
indeed quite challenging. An alternative approach would be to perform a systematic large coupling
expansion of the quantities treated in this work, which could perhaps lead to the time derivatives
of all the moments. We remind that  in the case of $K=1$ the leading term of such an expansion was already given in
\cite{jerome-atom-loss}. Our results could be used as a benchmark for such a large coupling
expansion: It is relatively easy to expand the multiple integrals into powers of $1/c$, in order to
facilitate future comparisons.

It would be interesting to consider the attractive Lieb-Liniger model as well.
In that model the fundamental particles can form bound states of arbitrary size. This might seem
like a serious complication as opposed to the repulsive case, however, certain properties of that
model are actually simpler. We expect to find relatively simple string-charge relations and also
asymptotic inverse operators for the transfer matrices. We note that already for quantum quenches it was found
that the attractive 
case displays simpler properties: for example in the quench from the Bose-Einstein
condensate (BEC) state the root densities were found to be polynomials
\cite{lorenzo-pasquale-attractive-LL,lorenzo-pasquale-attractive-LL2} as opposed to the special functions that appear in the
repulsive case \cite{caux-stb-LL-BEC-quench}. 

It is desirable to compare our present results to those of
\cite{alvise-lorenzo-1,alvise-lorenzo-2}, which treated the 
local correlations of the 1D Bose gas using a non-relativistic limit of the sinh-Gordon model.
There the $K$-body
correlator is expressed using a different family of auxiliary functions, such that a practical and closed form
result is found for all $K$. Thus no factorization was needed in
\cite{alvise-lorenzo-1,alvise-lorenzo-2}, and this is a considerable advantage over the earlier
approaches. If the methods of \cite{alvise-lorenzo-1,alvise-lorenzo-2} could be extended to
treat the particle losses, then this could perhaps lead to more compact exact formulas.

Finally, it would be interesting to compare our formulas to the numerical results of
\cite{jerome-atom-loss}.

We hope to return to these open questions in a future work.

\begin{acknowledgments}
We would like to thank Jer\^ome Dubail and Andrii Liashyk for useful discussions. 
\end{acknowledgments}

\appendix

\section{Similarity transformation}

\label{sec:similarity}

The existing literature dealing with scalar products and correlation functions, and in particular
Slavnov's seminal work \cite{slavnov-overlaps} uses an $R$-matrix, which is not identical to
\eqref{R}, and which is given by 
\begin{equation}
\label{R2}
\tilde R(u-v)=
\begin{pmatrix}
f(u,v) & 0 & 0 & 0 \\
0 &  1 & g(u,v) & 0 \\
0 & g(u,v) & 1  &  0 \\
0 & 0 & 0 & f(u,v) \\
\end{pmatrix},
\end{equation}
up to multiplicative normalization. It can be seen that the difference between \eqref{R} and
\eqref{R2} is {\it not} a similarity (or gauge) transformation. Instead it is given by a linear
transformation
\begin{equation}
  \label{Rdefdef}
\tilde R_{1,2}(u)=G_1 G_2^{-1} R_{1,2}(u) G_1 G_2^{-1},
\end{equation}
where $G_j$ with $j=1,2$ is a linear operator given for both spaces by
\be{Sdeform}
G=q^{S_z/2}=
\begin{pmatrix}
q^{1/4} & \\
& q^{-1/4}
\end{pmatrix}.
\ee

Note that both $R(u)$  and $\tilde R(u)$ preserve the $U(1)$ charge, thus
\begin{equation}
  R_{12}(u) G_1G_2=G_1G_2 R_{12}(u),
\end{equation}
and similarly for $\tilde R(u)$. It can be shown using this identity that the Yang-Baxter relation \eqref{YB2}
holds for $R(u)$ if and only if it holds for $\tilde R(u)$. The computation consists of mere
substitution and commuting through the appropriate factors.

In the main text it is stated that the Lax operator \eqref{eredetiL} satisfies the RLL relation
\eqref{RLL} with the $R$-matrix \eqref{R}. This exchange relation can be checked by direct
computation.

Similar to the relation \eqref{Rdefdef} let us also construct a new Lax operator. Let
\be{tildeL}
  \tilde L_j(u)=G_a q^{N_j/2} L_j(u) G_a q^{N_j/2}.
\ee
The Lax operator also conserves the $U(1)$-charge:
\be{commute}
  L_j G_a q^{ -N_j/2}=  G_a q^{ -N_j/2} L_j.
\ee
Using this property it can be shown that the $\tilde L_j(u)$ operators satisfy the RLL relation
\eqref{RLL} with the $R$-matrix \eqref{R2}.

An important observation is that in \eqref{tildeL} the effect of the $G_a$ factors is easily
canceled with a shift in the rapidity parameter:
\begin{equation}
  \label{tildeL2}
  \tilde L_j(u-\eta/2)= q^{N_j/2} L_j(u) q^{N_j/2}.
\end{equation}
The $R$-matrix depends only on the rapidity differences, thus the operators $\tilde L_j(u-\eta/2)$
also satisfy the RLL relation with the $R$-matrix \eqref{R2}.

Let us now construct a monodromy matrix using the deformed and shifted Lax operators:
\begin{equation}
  \tilde  T(\lambda)=\tilde L_L(\lambda-\eta/2) \dots \tilde L_1(\lambda-\eta/2).
\end{equation}
Making use of \eqref{tildeL2} it is seen that 
\begin{equation}
  \label{Trel}
   \tilde  T(\lambda)=q^{N/2} T(\lambda) q^{N/2},
\end{equation}
where $T(\lambda)$ is the monodromy matrix defined in the main text in \eqref{eq:T} and
\begin{equation}
  N=\sum_{j=1}^L N_j
\end{equation}
and we used that the $N_j$ commute with the q-boson operators at site $k$ for every $k\ne j$. 

The relation \eqref{Trel} holds for every matrix element of the monodromy matrix separately.
These relations can be used to relate the scalar products computed with the two different
conventions. It is easy to see that the proportionality factors between Bethe states depend only on
the particle number and not the rapidities, thus for any two sets $\bar\lambda$ and $\bar \mu$ we have
\begin{equation}
  \frac{\bra{0}C(\bar\lambda)B(\bar\mu)\ket{0}}{\bra{0}C(\bar\mu)B(\bar\mu)\ket{0}}=
 \frac{\bra{0}\tilde C(\bar\lambda)\tilde B(\bar\mu)\ket{0}}{\bra{0}\tilde C(\bar\mu)\tilde B(\bar\mu)\ket{0}}.
\end{equation}
This is guaranteed by the relations \eqref{Trel}.

The ratio of scalar products on the r.h.s. above can be computed with the Slavnov determinant, because
the operators involved satisfy the exchange relations given by the $R$-matrix \eqref{R2}. This is
the $R$ matrix which was
used in \cite{slavnov-overlaps}. On the other hand, the equality tells us that we can use the same
formulas also in our construction, given that we consider such normalized scalar products.

\section{Action of $\qB$ and $\qB_1^{\dagger}$\label{homogeneous1}}

\subsection{Homogeneous limit for the action of $\qB_1$\label{homogeneous2}}

Using {the standard steps from \cite{Korepin-Book} we find} that the action of a single $C$ operator is given by
\begin{multline}
\label{C_action}
C(\xi)B(\bar\mu)|0\rangle=\\
=\sum_{k=1}^N\left(d(\xi)a(\mu_k)X_k+a(\xi)d(\mu_k)\tilde X_k\right)B(\bar\mu_k)|0\rangle\\
+\sum_{j<k}\left(a(\mu_j)d(\mu_k)Y_{jk}+d(\mu_j)a(\mu_k)Y_{kj}\right)\\
\times B(\{\bmu_{jk},\xi\})|0\rangle,
\end{multline}
where
\be{XX}
\begin{split}
&X_k=g(\xi,\mu_k)f(\bar\mu_k,\mu_k)f(\xi,\bar\mu_k)e^{-2\eta N+\eta},\\
&{\tilde X_k}=g(\mu_k,\xi)f(\mu_k,\bar\mu_k)f(\bar\mu_k,\xi)e^{-2\eta N+\eta},
\end{split}
\ee
and
\begin{multline}
\label{Y}
Y_{jk}=g(\xi_j,\mu_k)g(\mu_j,\xi)\\
\times f(\mu_k,\mu_j)f(\mu_k,\bar\mu_{jk})f(\bar\mu_{jk},\mu_j)e^{-2\eta N+\eta}.
\end{multline}
Using now 
\be{g_limit}
\begin{split}
&{\lim_{\xi\to -i\infty}   g(\mu,\xi)\sim -2e^{-|\xi|} e^{i\mu}i\sin(i\eta)},\\
&{\lim_{\xi\to -i\infty}   f(\mu,\xi)\sim e^{-\eta}},
\end{split}
\ee
it is easy to obtain from \eqref{C_action} and \eqref{XX}--\eqref{Y} the following action  of a
single operator $\psi_1$:
\be{B_action}
  \psi_1\ket{\{\mu\}}=\chi e^{-(N-1)\eta}\sum_j
a(\mu_j){e^{i\mu_j} } f(\mu_j,\bar \mu_j)
  \ket{\{\mu_j\}}.
\ee
The multiple action is found by iterative procedure. Let us consider the double action of $\psi$, then
\begin{multline}
\label{psi_2}
\psi_1^2|\{\mu\}\rangle=\chi^2e^{-(2N-3)\eta}\sum_ja(\mu_j){e^{i\mu_j} }f(\mu_j,\bmu_j)\\
\times\sum_{j\ne k}a(\mu_k){e^{i\mu_k} }f(\mu_k,\bmu_{jk})|\{\mu_{jk}\}\rangle.
\end{multline}
Let us rewrite the double sum in \eqref{psi_2} as a sum over partition of $\bu\to\{\bu_{\so},\bu_{\st}\}$ with $|\bu_{\so}|=2$, then
\begin{multline}
\label{psi_2_iteration}
\psi_1^2|\{\mu\}\rangle=\chi^2e^{-(2N-3)\eta}\sum_{\bu\to\{\bu_{\so},\bu_{\st}\}}a(\bmu_{\so}){e^{i\bmu_{\so}} }f(\bmu_{\so},\bmu_{\st})\\
\times\sum_{\bu_{\so}\to\{u_j,u_k\}}f(u_j,u_k)|\{\mu_{\st}\}\rangle.
\end{multline}
We can compute the last sum over partition explicitly. Indeed
\be{summation2}
\sum_{\bx_{\so}\to\{x_1,x_2\}}f(x_1,x_2)=e^{\eta}+e^{-\eta}.
\ee

Similarly, in a general case
\begin{multline}
\label{psi_3_iteration}
\psi_1^K|\bmu\rangle=\chi^2e^{-(2N-3)\eta}\sum_{\bmu\to\{\bmu_{\so},\bmu_{\st}\}}a(\bmu_{\so}){e^{i\bmu_{\so}}}f(\bmu_{\so},\bmu_{\st})\\
\times\sum_{\bmu_{\so}\to\{\bmu_{i},\bmu_{ii}\}}f(\bmu_i,\bmu_{ii})|\{\mu_{\st}\}\rangle,
\end{multline}
with $|\bmu_{i}|=1$, $|\bmu_{ii}|=K-1$ it can be checked  that 
\begin{multline}
\label{summationK}
\sum_{\bmu_{\so}\to\{\mu_i,\bmu_{ii}\}}f(\bmu_i,\bmu_{ii})
=\left(e^{K\eta}+\dots+e^{-K\eta}\right).
\end{multline}

Thus in the case of $\psi_1^K|\{\bmu\}\rangle$ we have a common factor
\begin{multline}
\label{multi_factor}
\left(\chi^K \prod_{s=1}^{K-1}e^{\eta s}\right)\frac{e^{-\eta NK}}{(2\sinh(\eta))^{K}}\prod_{\ell=1}^{K}\left(1-e^{-2\eta\ell}\right)=\\
=\frac{\chi^{K}}{(2\sinh(\eta))^K}e^{\eta(K(K+1)/2-NK)}\prod_{\ell=1}^{K}\left(1-e^{-2\eta\ell}\right).
\end{multline}
Here we used $2\sinh(\eta)\sum_{\ell=1}^ke^{-2\ell\eta}={e^{\eta}}\left(1-e^{-2k\eta}\right)$. Thus we arrive at \eqref{CC_red}.

\subsection{Scalar product \label{scalar_prod}}

Here we compute the scalar products relevant to the main computation. As explained in Appendix
\ref{sec:similarity}, the normalized scalar products can be obtained using the known formulas in the
literature that use the $R$-matrix as given by \eqref{R2}. Additional factors of $q$ which arise in
the un-normalized scalar products eventually cancel.

The scalar product of an eigenvector $\langle\{\muc\}|$ and an arbitrary vector $|\{\mub\}\rangle$ of form \eqref{Bstate} is given by \cite{slavnov-overlaps}
\begin{multline}
\label{scalp}
\langle \{\muc\}|\{\mub\}\rangle={d(\bmub)d(\bmuc)\Delta'_g(\bmuc)\Delta_g(\bmub)}\\
\times h(\bmub,\bmuc)\det \mathcal M(\{\mub\}|\{\muc\}),
\end{multline}
where {$d(v)=e^{-iLv}$, $a(v)=e^{iLv}$}, $r(v)=a(v)/d(v)$,  and the corresponding matrix is given by
\begin{multline}
\label{slavnov_matrix}
\mathcal M\left(\{\mub\}|\{\muc\}\right)\\
=t(\mub_k,\muc_j)+ t(\muc_j,\mub_k){r(\mub_k)}\frac{h(\bmuc,\mub_k)}{h(\mub_k,\bmuc)},\\
\qquad j=1,\dots,N,\qquad k=1,\dots,N.
\end{multline}
Here we also used the function
\be{t-def}
t(\mu,\lambda)=g(\mu,\lambda)/h(\mu,\lambda).
\ee
Note that in the limit $\bmuc\to\bmub$ the matrix $\mathcal M$ becomes identical to the Gaudin
matrix \eqref{Gaudin}. 

Using the definition \eqref{BCfields}  the action  of $\psi^{\dagger K}_1$ can be presented as
\be{BtoXi}
{\bra{\{\mu\}}\psi^{\dagger K}_1 \ket{\{\mu_{\st}\}}=\chi^{-K}\lim_{\bar\xi\to i\infty}\left[e^{i\bar\xi (L-1)}\langle\{\mu\}|\{\xi,\mu_{\st}\}\rangle\right].}
\ee
{Here $|\bmu_{\st}|+K=|\bmu|$.} We take the limits such that we first send $\xi_1\to\dots\xi_K\to\xi$ and afterwards we take $\xi\to i\infty$.  In the homogeneous limit $\xi_i\to\xi_j$ the determinant  of $\mathcal M$ becomes zero, while
the pre-factor becomes singular. A finite limit is reached, whose value needs separate treatment.

It is clear that only the first term in each of the first $K$ columns of \eqref{slavnov_matrix} will survive, since the second term contains the factor $r(\xi)=e^{2i L\xi}\to 0$ at $\xi\to i\infty$. Rewrite the
first $K$ columns of $\mathcal M$ as the series w.r.t. $z=e^{|\xi|}$
\begin{multline}
\label{M_q-expand}
{t(\xi_p,\mu_j)}=\frac{{4 i^2\sin^2(i\eta)}}{z_p^2a_jb_j\left(1-(z_pa_j)^{-2}\right)\left(1-(z_pb_j)^{-2}\right)}\\
={i^2\sin^2(i\eta)}\frac{4}{z_p^2a_jb_j}\sum_{\ell,n}(a_jz_p)^{-2\ell}(b_jz_p)^{-2n}.
\end{multline}
Here we denote {$z_p=e^{-i\xi_p}$, $a_j=e^{i\mu_j}$ and $b_j=e^{i\mu_j+\eta}$}. We expand
\eqref{M_q-expand} in the Taylor series w.r.t. $z_p^2$ around some point $z$
\begin{multline}
\label{Taylor_2q}
{4i^2\sin^2(i\eta)}\sum_k\frac{1}{k!}(z_p^2-z^2)^k\\
\times\frac{\partial^k}{\partial(z_p^2)^k}\left[\sum_{\ell,n}a_j^{-2\ell-1}b_j^{-2n-1}z_p^{-2(n+l+1)}\right].
\end{multline}
In order to keep the columns of determinant linearly independent  we take in the first $K$ columns
and extract factor $(z^2_p-z^2)^k z_p^{-2k}$ ($k=1,\dots,K$) from the each column. This factor in
the limit $z_1\to z_2\to\dots\to z_K\to\infty$ cancels with the factor $\Delta_g(\bar z)$ such that
\be{cancelation_1}
\prod_{k=1}^K(z^2-z_p^2)^k/z_p^{2k}\:\Delta_g(\bz)=\left(2{i\sin(i\eta)}\right)^{K(K-1)/2}.
\ee

 The limit in the determinant can be also taken. Extracting form each column $z_p^{2(\ell+n+1)}$ and
 taking the homogeneous limit $z_1\to z_2\to\dots\to z_K\to z$ (and further $z=e^{\xi}\to \infty$)
 we get that the total order is $\xi^{-K(K+1)}$. We have in each column $a_j^{-2k-2}{e^{-(2n+1)\eta}
 }$, $k=\ell+n$, and a different power of $k$ should be taken in each column otherwise the columns will be dependent again (in the $k$-th column we keep only $k$-th power of $e^{\mu_j}$ since all lower powers can be canceled by subtraction of linear combination of different columns), thus we need to take in the first column  $\ell=n=0$, in the second $\ell+n=1$, in the third $n+\ell=2$, etc. Then the determinant can be written as
\begin{multline}
\label{M_q-limit}
\det \mathcal M\left(\{\mu_{\st},\xi\}|\{\mu\}\right)
=\det \tilde M\left(\{\mu_{\st}\}|\{\mu\}\right) e^{-\xi K(K+1)},\\
 k=1,\dots,K,
\end{multline}
with matrix elements of $\tilde M\left(\{\mu_{\so}\}|\{\mu\}\right)$ are given by
\be{tildeM}
\begin{split}
\tilde M_{jk}&=C_ke^{{-2i k \mu_j}}{\sum_{\ell=1}^k}e^{-2\ell \eta},\\
\tilde M_{jk}&=t(\mu_p,\mu_j)+ t(\mu_j,\mu_p){r(\mu_p)}\frac{h(\bmu,\mu_p)}{h(\mu_p,\bmu)},
\end{split}
\ee
where $C_k=4e^{-\eta}(-1)^{k-1}{i^2\sin^2(i\eta)}$.

On the other hand, the pre-factor of the determinant gives at the corresponding order $z^{K^2}$
\begin{multline}
\label{lim1}
h(\bmu_{\so},\bar\xi)h(\bmu_{\st},\bar\xi)g(\bmu_{\st},\bar\xi)\to h(\bmu_{\so},\bar\xi){e^{\eta NK}}\\
\to {(-z)^{K^2}}{\left(2i\sin(i\eta)\right)}^{-K^2}e^{\eta KN}{e^{iK\bmu_{\so}}.}
\end{multline}
Thus collecting all powers of $z=e^{\xi}$ from \eqref{M_q-limit}, \eqref{lim1}, extracting  factors $C_k$ from the first $K$ columns of the determinant  and taking into account the additional power of {$\sin(i\eta)$} from \eqref{cancelation_1} we obtain the following behavior of the scalar product:
\begin{multline}
\label{scal_q-limit}
{\lim_{\bar\xi\to i\infty}}\langle\{\mu\}|\{\xi,\mu_{\st}\}\rangle={e^{iK\bmu_{\so}} }\prod_{k=1}^K\left(1-e^{-2k\eta}\right)\\
\times\frac{e^{\eta KN}}{\left(2\sinh(\eta)\right)^{K(K-1)/2}}d(\bmu)d(\bmu_{\st})\Delta_g(\bmu)\Delta_g(\bmu_{\st})\\
\times h(\bmu,\bmu_{\st})\det\tilde{\mathcal M}\left(\{\mu_{\st}\}|\{\mu\}\right)\cdot d(\bar\xi)e^{-\xi K},
\end{multline}
where
\be{Mtilde}
\begin{split}
\tilde{\mathcal M}_{jk}&={e^{-2ik\mu_j}},\qquad k=1,\dots,K,\\
\tilde{\mathcal M}_{jk}&=\mG_{jk},\qquad k=K+1,\dots,N.
\end{split}
\ee
Above we also used the equality ${2\sinh(\eta)}\sum_{\ell=1}^ke^{-2\ell\eta}={e^{\eta}}(1-e^{-2k\eta})$.

Note now, that the definition \eqref{BCfields} can be written as 
\be{B_1d}
\qB_1^{\dagger}=\chi^{-1}{\lim\limits_{\xi\to i\infty}}\left[\left(d^{-1}(\xi)\cdot B(\xi)\right)e^{\xi}\right],
\ee
thus factor $d(\bar\xi)e^{-\xi K}$ from \eqref{scal_q-limit} cancels  with the $\xi$-dependent
factors in \eqref{B_1d}. 
Hereby we observe the cancellation of all singular factors in the formula for the action of the field $\psi_1^{\dagger}$.

Thus, finally we arrive at
\begin{multline}
\label{B1R}
\langle\{\mu\}|\psi_1^{\dagger K}|\{\mu_{\st}\}\rangle=\chi^{-K}{e^{iK\bmu_{\so}}}\prod_{k=1}^K\left(1-e^{-2k\eta}\right)e^{\eta KN}\\
\times\frac{\Delta_g(\bmu)\Delta_g(\bmu_{\st})}{\left(2\sinh(\eta)\right)^{K(K-1)/2}}h(\bmu,\bmu_{\st})\det\tilde{\mathcal M}\left(\{\mu_{\st}\}|\{\mu\}\right).
\end{multline}

\section{Computation of the function $F$}

\label{sec:F}

Here we consider the function $F\left(\{ p\}\right)$ defined in \eqref{F_def_scale}.
It is easy to see that this function is in fact a constant. First note that $F(\{ p\})$ is a symmetric
function of the rapidities in $\bar p$. Thus in order to prove that $F\left(\{ p\}\right)$ is a constant
it is enough to prove that it does not depend on one of the variables. Let us consider the poles of
\eqref{F_def_scale}. The function has simple poles w.r.t. some fixed variable $p_k$ at points $p_1,
p_2,\dots$. Let us calculate residues at these poles. We rewrite $F\left(\{ p\}\right)$ in the following form 
\begin{multline}
\label{Residue}
F\left(\{p\}\right)\\=\sum
\left[f(\bar p_k,p_b)f(p_k,p_b)-f(p_b,p_k)f(p_b,\bar p_k)\right]2ip_b\\
+2i\left[f(\bar p,p_k)-f(p_k,\bar p)\right]p_k,
\end{multline}
where summation in the first term is taken over partitions $\bar p\to\{\bar p_{\so},p_b\}$, $p_b\ne p_k$ and we write separately the term with $p_b=p_k$.  We explicitly write in the first term $f(\bar p_{\so},x)=f(\bar p_k,x)f(p_k,x)$ where the notation $\bar p_k=\bar p_{\so}\setminus p_k$ is used. Taking the residue of $F\left(\{ p\}\right)$ at $p_k=p_s$ for arbitrary $s\ne k$ (note that in the sum in $F\left(\{ p\}\right)$ only one term with $p_b=p_s$ survives) we arrive at
\begin{multline}
\label{Res_cancel}
\Res_{p_k=p_s} F\left(\{ p\}\right)=2ic\left[f(\bar p_k,p_s)+f(p_s,\bar p_b)\right]p_s\\
-2ic\left[f(\bar p_k,p_s)+(p_s,\bar p_k)\right]p_s=0.
\end{multline}
Thus $F\left(\{ p\}\right)$ does not have a pole w.r.t. $p_k$ at $p_s$ for arbitrary $k\ne s$. Also it is easy
to see that residue of $F\left(\{ p\}\right)$ at infinity is  zero. Then we conclude that $F\left(\{ p\}\right)$ is a
constant.

This constant can be found by recursion. 
Let us denote by $F_K$ the value of this constant, which
refers to the evaluation on the set with $|\bar p|=K+1$. It can
be checked that the
initial condition is $F_1=4c$. For the value of $F_{K}$ let us take the $p_1\to\infty$ limit of
the formula. 
Let us separate the terms and write
\begin{multline}
  F_{K}=\lim_{p_1\to\infty} \sum_{j=2}^{K+1}
  2ip_j \Big( f(\bar p_{j,1},p_j) f(p_1,p_j)\\ 
- f(p_j,\bar p_{j,1})f(p_j,p_1)\Big)\\
+\lim_{p_1\to\infty}    2ip_1 \Big( f(\bar p_{1},p_1) -
  f(p_1,\bar p_{1})\Big).
\end{multline}
The first terms can be found using the limit
\be{limf}
  \lim_{u\to\pm\infty} f(u,x)=1.
\ee
The second term can be found by using the sub-leading contributions in the $f$ functions.
Finally we obtain the recursion
\be{recursion}
  F_{K}=F_{K-1}+4cK.
\ee
The solution of this equation with the given initial condition is $F_K=2K(K+1)c$.  

\section{Factorization\label{Factorization}}

Here we consider the factorization of multiple integrals given by equation \eqref{direct_integral}.
For convenience we omit here some of the pre-factors and factorize the multiple integrals of the
form
\be{factor_a}
J_K=\int (d\bar p)\
s(\{p\})
E(\{p\})
\prod_{j=1}^K h^{(j-1)}(p_j),
\ee
where the short notation for the integrals was defined in \eqref{barpint}. We will see that the factorized expressions become combinations of the quantities $\{n,m\}$ defined
in \eqref{nm}.

Before turning to the details, let us explain the key idea of the factorization, as it was first
laid out in \cite{sajat-XXZ-to-LL}. The goal is to
manipulate the rational functions in the integrand such that a certain degree of separation is
achieved. We intend to write the integrand as sums of terms of the type
\be{integrand}
  \hdots \frac{1}{(p_j-p_k)^2+c^2}h^{(l)}(p_k),
\ee
where $j$, $k$ and $l$ are arbitrary indices, and the dots stand for a product of further $h^{(m)}$ functions and an arbitrary
rational function of the rapidities, such that this product does not depend on $p_k$. In such a case
the integral over $p_k$ can be performed using the definition \eqref{hhdef}, by noting that
\be{reduction}
   \frac{1}{(p_j-p_k)^2+c^2}=\frac{1}{2c}\varphi(p_j-p_k).
\ee
After such a step the number of integrals (and thus the number of variables) is reduced by one. In
the next step further algebraic manipulations might be required, but the process can be repeated
with the same strategy. Eventually one ends up with the simple integrals given by \eqref{nm}.

At present there is no algebraic theory behind this procedure, and the computation need to be
performed on a case by case basis.

\subsection{K=1}

For $K=1$ we have a simple integral given by
\be{J1}
  J_1=\int dp \ f(p) p^2  h^{(0)}(p).
\ee
Using the definition \eqref{nm} this is equal to $c^3\{0,2\}$.

\subsection{K=2}

Consider now the case $K=2$:
\be{factor_1}
J_2=\int
(d\bar p)
\frac{(p_2-p_1)(p_1^2+p_2^2)}{(p_2-p_1)^2+c^2}
h^{(0)}(p_1) h^{(1)}(p_2).
\ee
We write the rational factors of the integrand as
\begin{multline}
\label{expand}
\frac{(p_2-p_1)(p_1^2+p_2^2)}{(p_2-p_1)^2+c^2}\\
=\frac{1}{3}\left\{\frac{2(p_2^3-p_1^3)+c^2(p_1-p_2)}{(p_1-p_2)^2+c^2}-(p_1-p_2)\right\}.
\end{multline}

Now we can perform the integrals separately, using the definitions of the auxiliary functions. After
the manipulations described above we obtain 
\begin{multline}
\label{factor_5}
  J_2=\frac{c^4}{6}\left(2(\{0,4\}-\{1,3\})  +\{1,1\}-\{0,2\}\right.\\
   \left. +2(\{0,0\}\{1,1\}-\{0,1\}^2)\right).
\end{multline}

\subsection{K=3}

Let us consider now the case $K=3$.
We use the following expansion
\begin{multline}
\label{factor_6}
\prod_{i<j}^3\frac{p_j-p_i}{(p_j-p_i)^2+c^2}\sum_{k=1}^3 p_k^2\\
=\sum_{\sigma\in\mathbb P}(-1)^{[\mathbb P]}D\left(p_{\sigma_1},p_{\sigma_2},p_{\sigma_3}\right),   
\end{multline}
where the sum is taken over the permutations of spectral parameters, $[\mathbb P]$ denotes the parity of  permutations and $D(x,y,z)$ is defined as
\be{Dxyz}
D(x,y,z)=\frac{z^3-z^2y+zc^2/3}{[(z-y)^2+c^2][(y-x)^2+c^2]}.
\ee
The equality \eqref{factor_6} can be checked by direct computation.

Applying this expansion we perform the integration over the first variable in each term:
\begin{multline}
\int(d\bar p)\frac{p_3^3-p_3^2p_2+c^2p_3/3}{[(p_3-p_2)^2+c^2][(p_2-p_1)^2+c^2]}\\
\hspace{1cm}\times h^{(0)}(p_1)h^{(1)}(p_2)h^{(2)}(p_3)+\perm.\\
=\frac{1}{2c}\int(d\bar p)\frac{p_3^3-p_3^2p_2+c^2p_3/3}{(p_3-p_2)^2+c^2}\\
\hspace{0.3cm}\times\left(h^{(0)}(p_2)-1\right)h^{(1)}(p_2)h^{(2)}(p_3)+\perm.      
\end{multline}

In the second line $\perm$ stand for analogous terms, with the appropriate replacements.

It can be seen that the terms proportional to $h^{(0)}(p_i)h^{(1)}(p_j)h^{(2)}(p_k)$ for
$i,j,k=1\dots3$ disappear, and the remaining terms are
\be{J3osszeg}
  J_3=J_3^{(a)}+J_3^{(b)}+J_3^{(c)}
\ee
with
\begin{multline}
\label{J3a}
J_3^{(a)}=-\frac{1}{2c}\int(d\bar p)\ h^{(1)}(p_2)h^{(2)}(p_3)\\
\times\frac{(p_3^3-p_2^3)+(p_2^2p_3-p_3^2p_2)+c^2(p_3-p_2)/3}{(p_3-p_2)^2+c^2},
\end{multline}
\begin{multline}
\label{J3b}
      J_3^{(b)}=\frac{1}{2c}\int(d\bar p)\ h^{(0)}(p_1)h^{(1)}(p_2) \\
\times \frac{2(p_2^2p_1^3-p_1^2p_2^3)+c^2(p_1p_2^2-p_1^2p_2)/3}{(p_1-p_2)^2+c^2},
\end{multline}
and
\begin{multline}
\label{J3c}
J_3^{(c)}=\frac{1}{2c}\int(d\bar p)\frac{p_1p_3^3-p_1^3p_3}{(p_1-p_3)^2+c^2}h^{(0)}(p_1)h^{(2)}(p_3).
\end{multline}

Now the problem of the factorization of \eqref{J3a}--\eqref{J3c} is reduced to the factorization of integrals
\be{xyz}
K^{(sr)}_{\alpha\beta}=
\int dxdy\frac{x^{\alpha}y^{\beta}-x^{\beta}y^{\alpha}}{(x-y)^2+c^2}
\frac{h^{(s)}(x)h^{(r)}(y)}{c^{\alpha+\beta+s+r}},
\ee
with $\alpha,\beta=1\dots4$, $r,s=0\dots 2$. Note that we introduced factors of $c$ in the
denominator so that each $K^{(sr)}_{\alpha\beta}$ is dimensionless.

Regarding $J^{(a,b,c)}$ we have
\begin{multline}
\label{J3a2}
J_3^{(a)}=\frac{c^5}{4}\Big[2K_{12}^{(23)}+\{2,4\}-\{1,5\}+\frac{1}{3}\left(\{2,2\}-\{1,3\}\right)\Big],\\
J_3^{(b)}=\frac{c^5}{2}\left[\frac{1}{3}K_{12}^{(12)}-2K_{23}^{(12)}\right],\hphantom{111111111111111111.}\\
J_3^{(c)}=\frac{c^5}{2}K_{13}^{(13)}.\hphantom{11111111111111111111111111.}
\end{multline}

For the rational functions appearing in the double integrals
above we perform the following manipulations:
\begin{multline}
      \frac{xy^2-x^2y}{(x-y)^2+1}
    =\frac{1}{3} (x-y)+\frac{1}{3}\frac{(y-x)+ (y^3-x^3)}{(x-y)^2+1},
\end{multline}
\begin{multline}
\frac{xy^3-x^3y}{(x-y)^2+1}
=\frac{1}{2}(x^2-y^2)+ \frac{1}{2}\frac{  (y^2-x^2)+    (y^4-x^4)   }{(x-y)^2+1},
\end{multline}
\begin{multline}
      \frac{x^2y^3-x^3y^2}{(x-y)^2+1}\\
=2/15 (x-y)+      2/5 (x^2 y-y^2 x)+      1/5 (x^3-y^3 )\\
+     \frac{   2/15 (y-x)+      1/3 (y^3-x^3)+      1/5 (y^5-x^5)  }{(x-y)^2+1}.
\end{multline}
Here the variable $x$ is be understood as the dimensionless combination $x=p/c$, with some $p$, and
similarly for $y$.

This separation of terms leads to the results
\begin{multline}
  \label{manyKa}  
K_{12}^{(12)}=
\frac{1}{6} ( \{0, 4\} -\{1, 3\} + \{0, 2\} - \{1, 1\} \\ 
+2 \{0, 1\}^2 -\{0, 0\} \{1, 1\}), 
\end{multline}
\begin{multline}
K_{12}^{(23)}=\frac{1}{6}\big(\{2,4\}-\{1,5\}+\{2,2\}-\{1,3\}\\
+2\left(\{2,1\}\{1,0\}-\{1,1\}\{2,0\}\right)\big),  
\end{multline}
\begin{multline}
     K_{13}^{(13)}=\frac{1}{4}\Big(2\left(\{0,2\}^2-\{2,2\} \{0,0\}\right)+ \{0,4\}\\
-\{2,2\} +\{0,6\}-\{2,4\}\Big),
\end{multline}
and
\begin{multline}\label{manyKd}
  K_{23}^{(12)}=\frac{1}{30}\Big( 4\left(\{0,1\}^2-\{0,0\}\{1,1\}\right)+3\{0,6\}\\
-3\{1,5\}+12\left(\{0,2\}\{1,1\}-\{0,1\}\{1,2\}\right)\\
+2\left(\{0,2\}-\{1,1\}\right)
+5\left(\{0,4\}-\{1,3\}\right)\\
+6(\{0,3\}\{0,1\}-\{1,3\}\{0,0\})\Big).
\end{multline}
Substituting the equations \eqref{manyKa}-\eqref{manyKd} to \eqref{J3a2} and finally into \eqref{J3osszeg} gives
  the final answer for $J_3$. The direct term for the energy loss is given by $\dire_{E,3}=(3!)^2
  J_3$ and this leads to formula \eqref{Yfin}.
  %

%

\end{document}